\documentclass{article}
\usepackage{graphicx} %For Including figures
\usepackage{authblk}
\usepackage{mwe} %For the blank example image
\usepackage{blindtext} %For the lorem ipsum text
\usepackage{siunitx} %Si-enheder
\usepackage{mathtools} %Lader os skrive formler
\usepackage{chemformula} %Tillader os at skrive kemiske formler
\usepackage[utf8]{inputenc}
\usepackage[style=phys,sorting=none,backend=biber]{biblatex} %For at stille fodnoter og referencer op på en pæn måde %Til at lave fodnoter og referencer
\usepackage{amsmath,amssymb}
\usepackage{bm}% boldmath
\usepackage{listings} % Code block (source code) \begin{lstlisting} 
\usepackage{graphicx}
\usepackage{lmodern}
\usepackage[textwidth=16cm,textheight=23cm]{geometry}

\usepackage{url}
\usepackage[colorlinks=true, pdfstartview=FitV, linkcolor=blue, citecolor=blue, urlcolor=blue]{hyperref}

\usepackage[font=small, labelfont=bf]{caption}
\usepackage{subfig} % Subfigures. Uses \subfloat[captions text]{figure}
\usepackage{float}

\usepackage{booktabs}   % Allows the use of \toprule, \midrule and \bottomrule in tables for horizontal lines
\usepackage{longtable}
\usepackage{array}
\usepackage{threeparttablex}

\usepackage{pgfplots}
\pgfplotsset{width=5cm,compat=1.9}

\usepgfplotslibrary{external}
\newcolumntype{x}[1]{%
>{\centering\hspace{0pt}}p{#1}}%

\usepackage{enumitem}

\usepackage{pstricks,pst-node,pst-tree} % includes graph additions
\usepackage{pst-pdf} % Compiles the pictures
\usepackage{pst-node}
\usepackage{pst-plot}
\usepackage{pst-3dplot}

\definecolor{lbcolor}{rgb}{0.9,0.9,0.9}  

\author[1]{Juliane H. Fuglsbjerg}
\author[1]{Dániel Nagy}
\author[2]{Hans Jørgen Aa. Jensen}
\author[1]{Stephan P. A. Sauer}
\affil[1]{\textit{Department of Chemistry, University of Copenhagen, Copenhagen Ø, Denmark}} \affil[2]{\textit{Department of Physics, Chemistry and Pharmacy, University of Southern Denmark, Odense M, Denmark}}
\date{}
\title{Performance of range-separated long-range SOPPA short-range density functional theory method for vertical excitation energies}

\begin{document}
\maketitle
\begin{abstract} 
\textbf{In this paper benchmark results are presented on the calculation of vertical electronic excitation energies using a long-range second-order polarisation propagator approximation (SOPPA) description with a short-range density functional theory (srDFT) description based on the Perdew-Burke-Ernzerhof (PBE) functional. The excitation energies are investigated for 132 singlet states and 71 triplet states across 28 medium sized organic molecules. The results show that overall SOPPA-srPBE always performs better than PBE, and that SOPPA-srPBE performs better than SOPPA for singlet states, but slightly worse than SOPPA for triplet states when CC3 results are the reference values.}
\end{abstract}

\textbf{Keywords:} benchmarks; electronically excited states; range-separation; second-order polarisation propagator method; SOPPA; SOPPA-srDFT; DFT; PBE

\section{Introduction}
Energy absorption in the range of electronic excitation energies is important within the field of chemistry. They can play a role in determining the energetics of chemical reactions and are fundamental for understanding how molecules interact with light. Electronic excitation energies are also important in several industrial applications such as dye-sensitised solar cells and artificial photosynthesis \cite{photo1,photo2,photo3}.
Here computational chemistry has great potential utilisation, predicting and fine-tuning chromophores and building blocks for new novel materials with particular excitation energies. However, in order to use computational chemistry for designing new materials with specific excitation energies, it is necessary to have computational methods that accurately predict electronic excitation energies while maintaining a low computational cost.

In this area time-dependent density functional theory (TD-DFT) \cite{TDDFT1,TDDFT2,TDDFT3,Dreuw2005} has had great success due to its low computational cost. However TD-DFT also has known shortcomings, for example with certain excitation classes such as charge transfer (CT) excitations which are often severally underestimated with TD-DFT \cite{CT1,CT2,CT3,CT4}. It has been shown for CT excitations that the underestimation arises because the regular exchange-correlation functionals behave incorrectly in regards to the distance between the charges \cite{CTdist1,CTdist2}. Attempts to meet this shortcoming have been made with range-separated hybrid functionals \cite{range1,range2,range3}, which separate the electron-electron repulsion term in a long- and a short-range part when calculating the exchange-correlation energy. CAM-B3LYP \cite{camb3lyp} is an example of such a functional.

These range-seperated hybrid functionals still exhibit significant shortcomings, for instance with excitations which have a high degree double excitation character \cite{DE1,DE2,DE3,DE4}. Here some wave function theory (WFT) methods such as multi-configuration self-consistent field (MC-SCF) are more suitable. An alternative approach could therefore be to combine DFT with WFT, so instead of correcting a short-range DFT description with a different long-range DFT description, it gets corrected with a long-range description using WFT. This long-range WFT short-range DFT (WFT-srDFT) approach have been developed for several WFT methods resulting in new methods such as MP2-srDFT \cite{MP2sr}, CI-srDFT \cite{CIsr}, CC-srDFT \cite{CCsr} and MC-srDFT \cite{MCsr1,MCsr2}.

In a previous paper \cite{soppasr} Hedegård $et$ $al.$ developed a method utilising a second-order polarisation propagator approximation (SOPPA) to describe the long-range interactions with a short-range description using a DFT functional based on the Perdew-Burke-Ernzerhof (PBE) functional \cite{CCsr}. SOPPA-srPBE showed promising results, performing better than regular SOPPA for both local and charge transfer excitations \cite{soppasr}. However the method was only tested for a small number of molecules and excitation states and the main focus of the study was charge transfer excitations.

In this paper a more thorough benchmarking of SOPPA-srPBE will be carried out to investigate the performance of SOPPA-srPBE compared to SOPPA and PBE when calculating vertical singlet and triplet excitation energies. SOPPA-srPBE will be tested using the Thiel benchmarking set \cite{cc32008}, which has also been used in previous benchmarking studies of other methods \cite{dft2008,noneiterative,cc32010,basiseffectCAS,geom1,geom2,cc32014,soppa,spas186}. The benchmarking set consists of 28 medium sized organic molecules and is intended to cover the most important chromophores in organic photochemistry \cite{cc32008}. 132 singlet excitation states and 71 triplet excited states will be computed using SOPPA-srPBE and PBE. The result will be benchmarked against SOPPA and CC3 results from previous studies \cite{soppa,cc32008,cc32010,cc32014} with the same geometries and basis set. Additionally the effect of adding extra diffuse basis functions to the basis set will be investigated for SOPPA-srPBE and compared to previous studies.

\section{Computational details}
All calculations are carried out using the DALTON program \cite{dalton}. The geometries employed are the same as in previous studies \cite{cc32008,dft2008,noneiterative,cc32010,basiseffectCAS,geom1,geom2,cc32014,soppa,spas186} and are optimised at the MP2/6-31G* level. Vertical singlet and triplet excitation energies were calculated using the SOPPA-srPBE method \cite{soppasr} and the Perdew-Burke-Ernzerhof (PBE) functional \cite{PBE}. The calculations were carried out in the TZVP \cite{TZVP} basis set. Additional calculations in the aug-cc-pVTZ \cite{aVTZ1,aVTZ2} basis set were carried out for both the singlet and triplet excitations with the SOPPA-srPBE method and the singlet excitations with the PBE method. The $\mu$ parameter in the SOPPA-srPBE calculation controls the range separation and was set to $\mu=0.4$, since this has been found in previous similar studies to be optimal \cite{mu1,mu2}.

The results will be compared to results from previous studies carried out with the same set of benchmarking molecules, geometries and basis sets. Specifically will the singlet excitation energies in both basis sets, and the triplet excitation energies in the TZVP basis set, be compared to the regular SOPPA results \cite{soppa} with CC3 results \cite{cc32008,cc32010,cc32014} as reference values in both the TZVP and aug-cc-pVTZ basis sets. %[some of the CC3 results are not proper but with a correction from another method].

\section{Results and discussion}
The benchmarking set consists of 28 medium sized organic molecules with 132 singlet excited states and 71 triplet excited states. In the following, the computed SOPPA-srPBE excitation energies  will be compared to PBE, SOPPA and CC3 results to investigate the performance of SOPPA-srPBE. The effect of adding the extra diffuse basis functions in the aug-cc-pVTZ basis set will also be investigated for SOPPA-srPBE. The excitation energies from this study along with the reference data are collected in tables \ref{tab:sing_tzvp}, \ref{tab:sing_aVTZ}, and \ref{tab:trip}. Statistical analyses of the results are collected in tables \ref{tab:sing_tzvp_stat}, \ref{tab:sing_avtz_stat}, \ref{tab:trip_stat}, and \ref{tab:stat_basis}.

\subsection{Singlet excitation energies with the TZVP basis set}
\begin{figure}[h!]
    \centering
    \subfloat{\includegraphics[width=5cm]{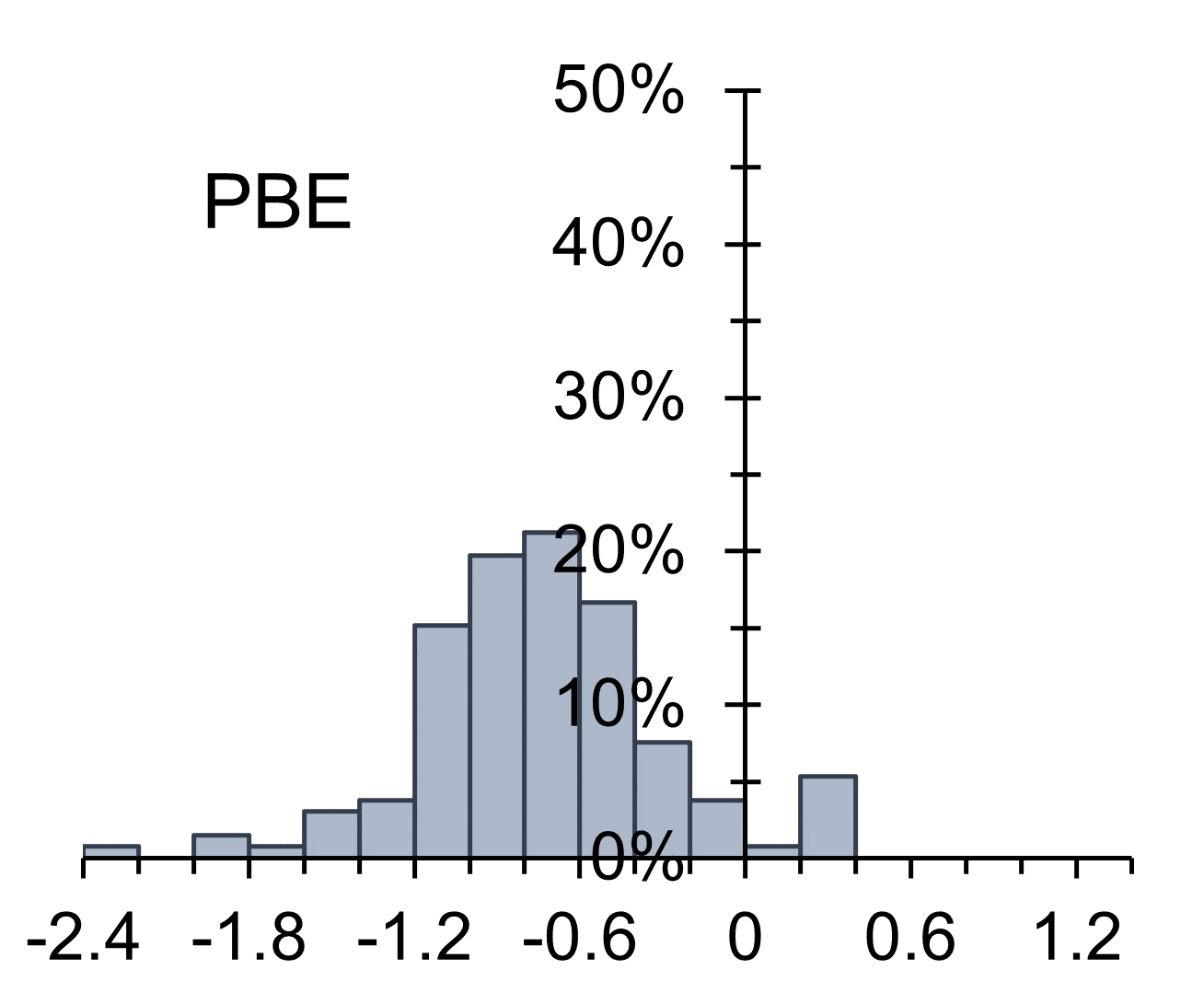}}
    \subfloat{\includegraphics[width=5cm]{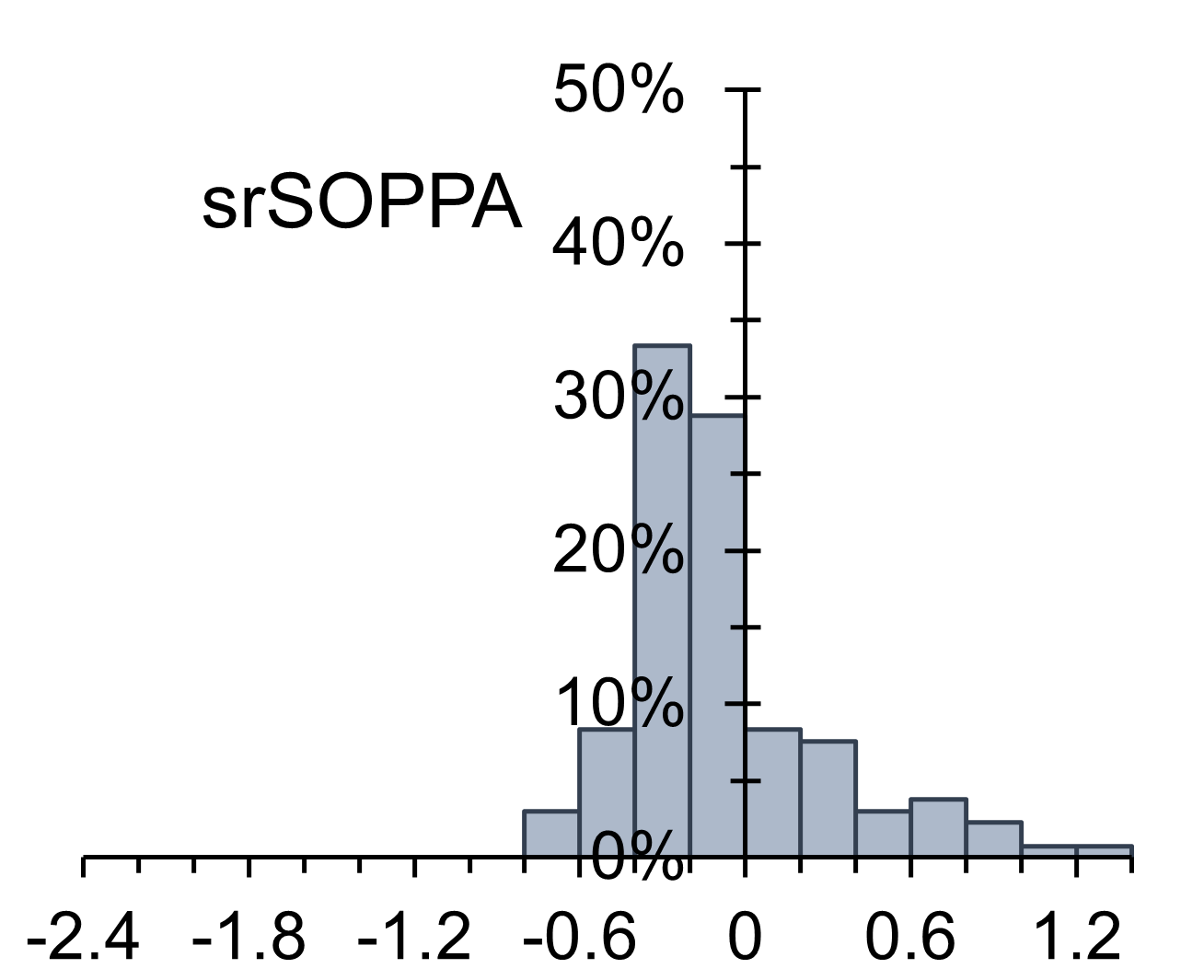}}
    \subfloat{\includegraphics[width=5cm]{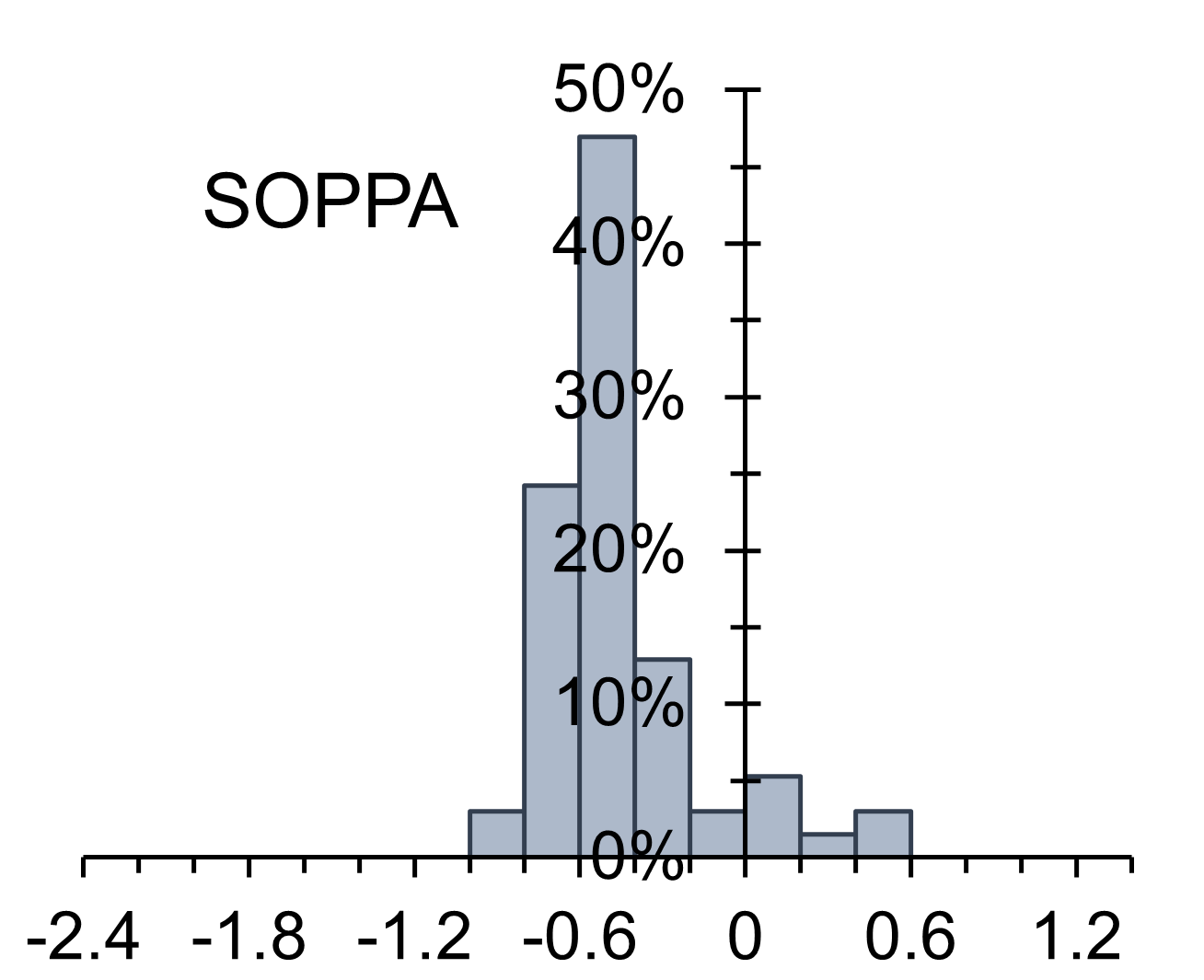}}
    \caption{Histograms (in \%) of the deviations of PBE, SOPPA-srPBE (srSOPPA) and SOPPA vertical singlet excitation energies (in eV) from the CC3 energies in the TZVP basis set.}
    \label{fig:sing_dev}
\end{figure}

\begin{figure}[h!]
    \centering
    \subfloat{\includegraphics[width=5cm]{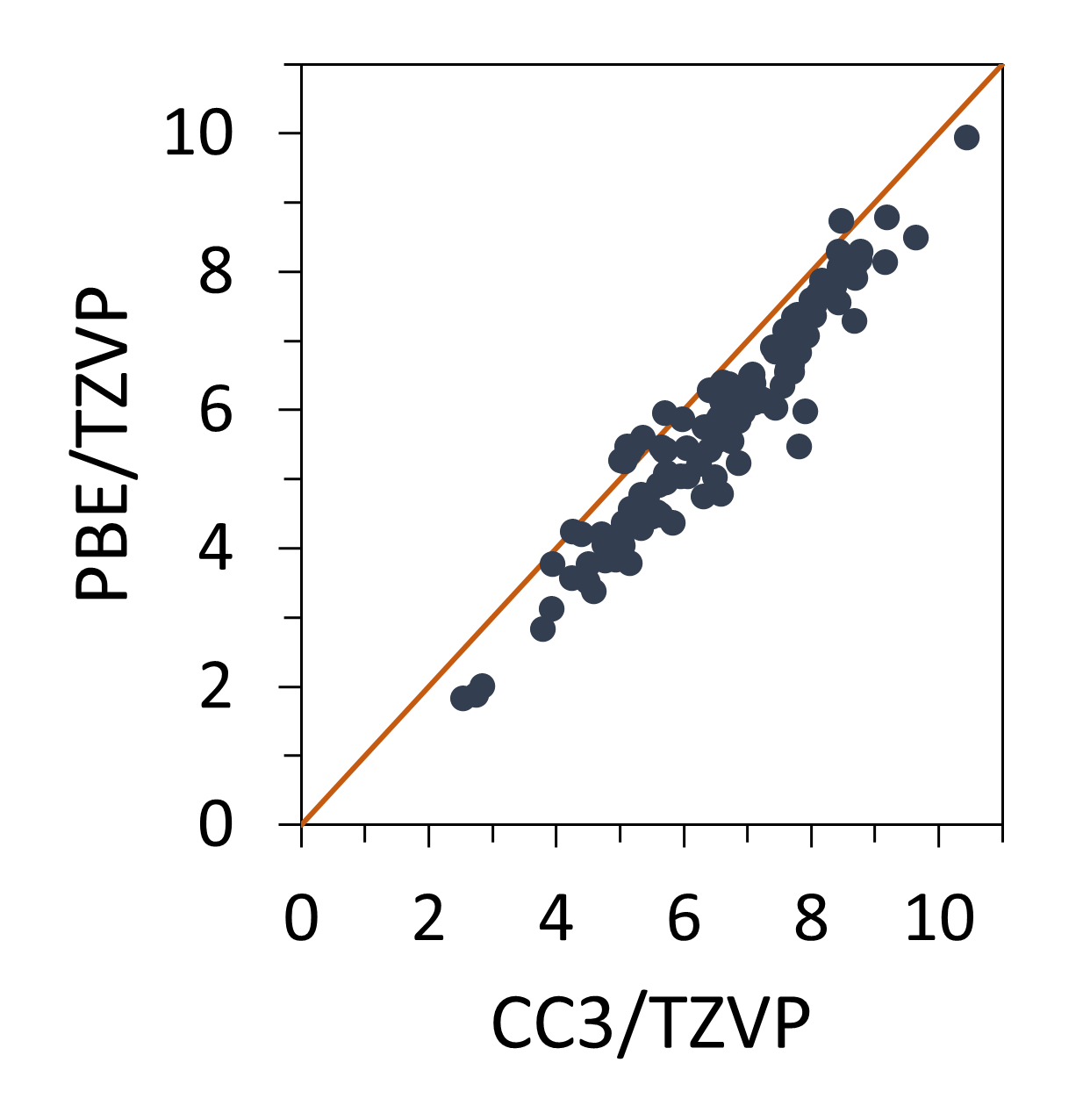}}
    \subfloat{\includegraphics[width=5cm]{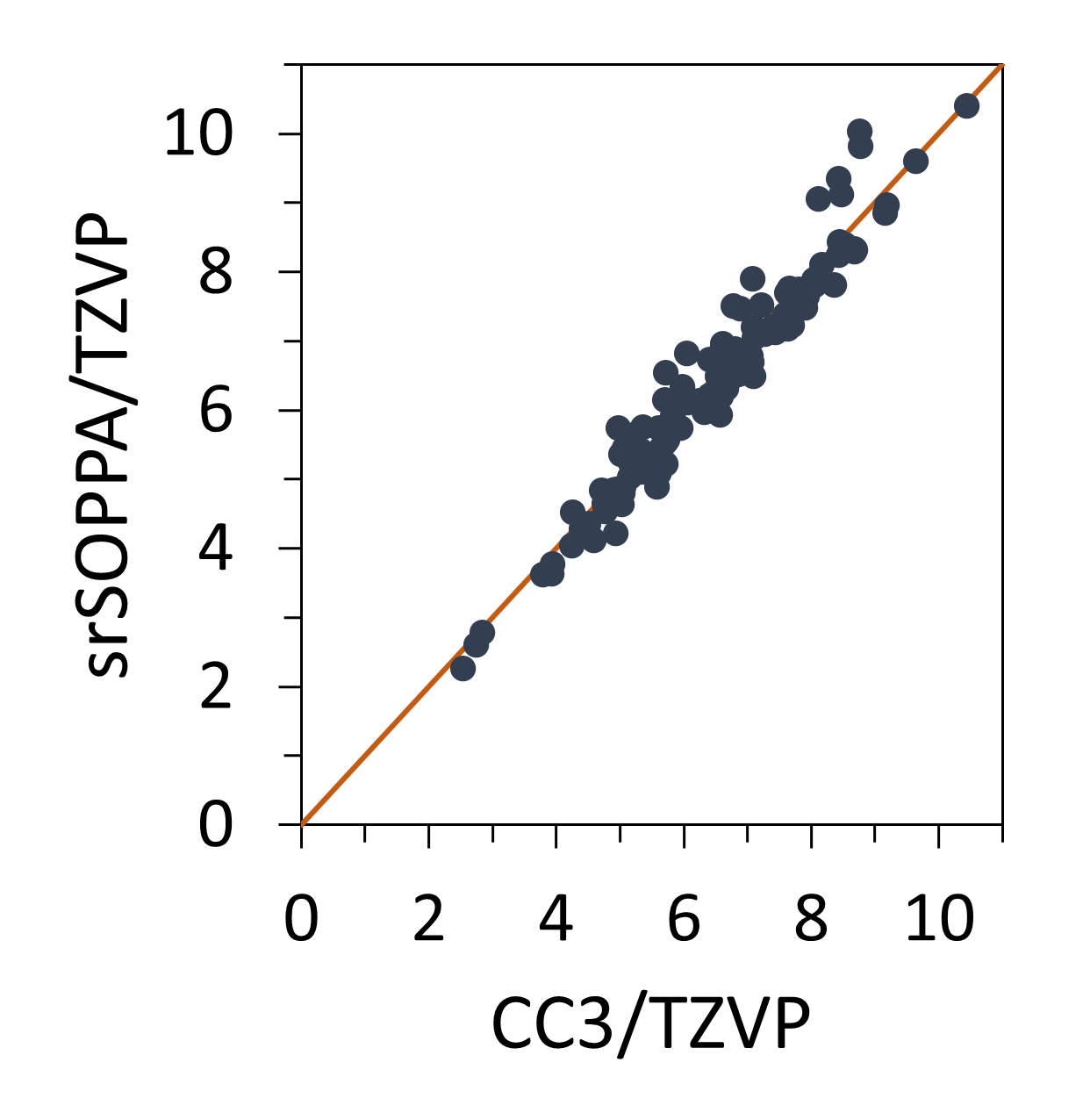}}
    \subfloat{\includegraphics[width=5cm]{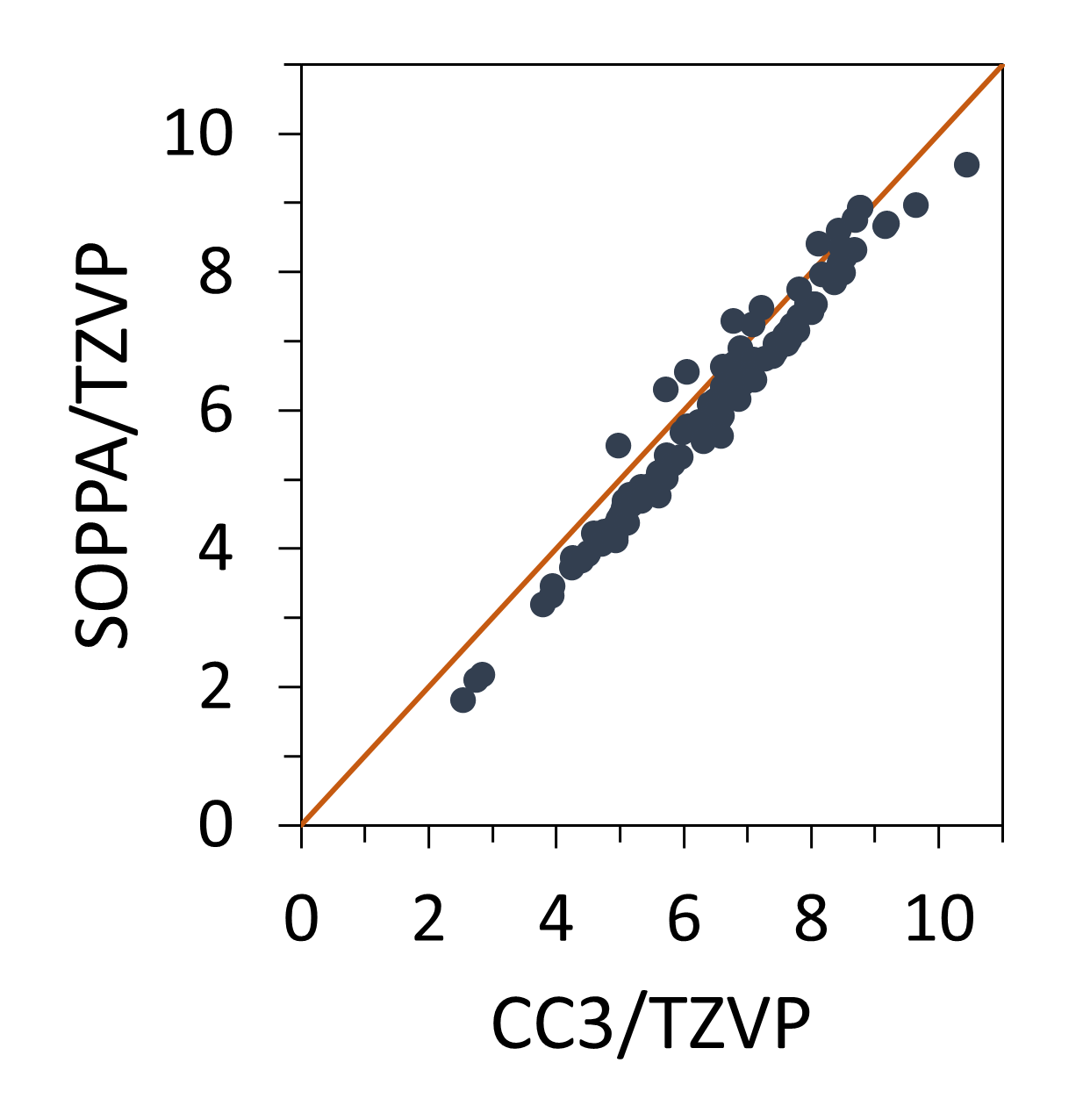}}
    \caption{Correlation plot of the vertical singlet excitation energies (in eV) with PBE, SOPPA-srPBE (srSOPPA) and SOPPA versus CC3 in the TZVP basis set.}
    \label{fig:sing_corr}
\end{figure}

In figure \ref{fig:sing_dev} histograms of the deviations from the CC3 results are shown for the PBE, SOPPA-srPBE and SOPPA results. In figure \ref{fig:sing_corr} the corresponding correlation plots can be seen. It can be seen from the histograms that SOPPA-srPBE both under- and overestimates CC3, whereas SOPPA and PBE primarily underestimate CC3 apart from a few outliers. The statistical analysis in table \ref{tab:sing_tzvp_stat} shows that SOPPA-srPBE has the lowest absolute mean deviation from CC3 ($0.30$ eV). This is an improvement from the mean deviations for SOPPA ($0.50$ eV) and PBE ($0.77$ eV). The change in absolute mean deviation is illustrated in \ref{fig:sing_dev} where the bulk of the histogram is the closest to $0$ eV for SOPPA-srPBE. For the standard deviations, the picture is slightly different. The standard deviation of SOPPA is the smallest ($0.28$ eV) followed by SOPPA-srPBE ($0.37$ eV) and lastly PBE ($0.45$ eV). By this measure SOPPA is performing the best of the three methods investigated. The standard deviations are reflected in figure \ref{fig:sing_dev} where the histograms get more spread out and in figure \ref{fig:sing_corr} where the points get more dispersed with increasing standard deviation.
\small
\begin{ThreePartTable}
\begin{TableNotes}
    \item[a] Results from \cite{soppa}
    \item[b] Results from \cite{cc32008} and \cite{cc32014}
\end{TableNotes}
\begin{longtable}[h!]{p{2.7cm}p{2.6cm}cccc}
\caption{Vertical singlet excitation energies (in eV) with the TZVP basis set.} \label{tab:sing_tzvp}

\\ \toprule Molecule & State & PBE & srSOPPA & SOPPA\tnote{a} & CC3\tnote{b} \\ \midrule \endfirsthead

\multicolumn{6}{l}{{\tablename} \thetable{}: \textit{continued.}} \\
\toprule Molecule & State & PBE & srSOPPA & SOPPA\tnote{a} & CC3\tnote{b} \\ \midrule\endhead

\bottomrule \multicolumn{6}{r}{\textit{continued.}} \endfoot 

\bottomrule \insertTableNotes \endlastfoot

Ethene & 1$^1$B$_{1u}$ ($\pi$ → $\pi^*$) & 7.78 & 7.80 & 7.84 & 8.37 \\[1.0ex]
E-Butadiene & 1$^1$B$_{u}$ ($\pi$ → $\pi^*$) & 5.62 & 5.93 & 5.88 & 6.58 \\
 & 2$^1$A$_{g}$ ($\pi$ → $\pi^*$) & 6.30 & 7.50 & 7.29 & 6.78 \\[1.0ex]
all-E-Hexatriene & 1$^1$B$_{u}$ ($\pi$ → $\pi^*$) & 4.52 & 4.88 & 4.80 & 5.58 \\
 & 2$^1$A$_{g}$ ($\pi$ → $\pi^*$) & 5.07 & 6.53 & 6.30 & 5.72 \\[1.0ex]
all-E-Octatetraene & 2$^1$A$_{g}$ ($\pi$ → $\pi^*$) & 4.19 & 5.73 & 5.49 & 4.98 \\
 & 1$^1$B$_{u}$ ($\pi$ → $\pi^*$) & 3.83 & 4.21 & 4.12 & 4.94 \\
 & 2$^1$B$_{u}$ ($\pi$ → $\pi^*$) & 5.44 & 6.82 & 6.55 & 6.06 \\
 & 3$^1$A$_{g}$ ($\pi$ → $\pi^*$) & 5.80 & 6.20 & 6.13 & 6.50 \\
 & 4$^1$A$_{g}$ ($\pi$ → $\pi^*$) & 5.97 & 6.88 & 6.68 & 6.81 \\
 & 3$^1$B$_{u}$ ($\pi$ → $\pi^*$) & 5.97 & 7.47 & 7.43 & 7.91 \\[1.0ex]
Cyclopropene      & 1$^1$B$_{1}$ ($\sigma$ → $\pi^*$) & 6.29 & 6.52 & 6.57 & 6.90 \\
 & 1$^1$B$_{2}$ ($\pi$ → $\pi^*$) & 6.13 & 6.48 & 6.65 & 7.10 \\[1.0ex]
Cyclopentadiene    & 1$^1$B$_{2}$ ($\pi$ → $\pi^*$) & 4.94 & 5.21 & 5.11 & 5.73 \\
 & 2$^1$A$_{1}$ ($\pi$ → $\pi^*$) & 6.09 & 6.91 & 6.63 & 6.62 \\
 & 3$^1$A$_{1}$ ($\pi$ → $\pi^*$) & 8.06 & 8.29 & 8.32 & 8.69 \\[1.0ex]
Norbornadiene      & 1$^1$A$_{2}$ ($\pi$ → $\pi^*$) & 4.48 & 5.14 & 5.05 & 5.64 \\
 & 1$^1$B$_{2}$ ($\pi$ → $\pi^*$) & 5.02 & 5.98 & 5.96 & 6.49 \\
 & 2$^1$B$_{2}$ ($\pi$ → $\pi^*$) & 6.60 & 7.17 & 7.12 & 7.64 \\
 & 2$^1$A$_{2}$ ($\pi$ → $\pi^*$) & 6.55 & 7.22 & 7.22 & 7.71 \\[1.0ex]
Benzene            & 1$^1$B$_{2u}$ ($\pi$ → $\pi^*$) & 5.25 & 5.45 & 4.69 & 5.07 \\
 & 1$^1$B$_{1u}$ ($\pi$ → $\pi^*$) & 6.03 & 6.31 & 6.15 & 6.68 \\
 & 1$^1$E$_{1u}$ ($\pi$ → $\pi^*$) & 6.03 & 7.14 & 6.96 & 7.45 \\
 & 2$^1$E$_{2g}$ ($\pi$ → $\pi^*$) & 8.29 & 9.34 & 8.6 & 8.43 \\[1.0ex]
Naphthalene        & 1$^1$B$_{3u}$ ($\pi$ → $\pi^*$) & 4.23 & 4.52 & 3.86 & 4.27 \\
 & 1$^1$B$_{2u}$ ($\pi$ → $\pi^*$) & 4.08 & 4.63 & 4.41 & 5.03 \\
 & 2$^1$A$_{g}$ ($\pi$ → $\pi^*$) & 5.86 & 6.33 & 5.68 & 5.98 \\
 & 1$^1$B$_{1g}$ ($\pi$ → $\pi^*$) & 5.04 & 6.11 & 5.77 & 6.07 \\
 & 2$^1$B$_{3u}$ ($\pi$ → $\pi^*$) & 5.74 & 5.96 & 5.74 & 6.33 \\
 & 2$^1$B$_{2u}$ ($\pi$ → $\pi^*$) & 5.89 & 6.33 & 6.08 & 6.57 \\
 & 2$^1$B$_{1g}$ ($\pi$ → $\pi^*$) & 6.19 & 6.49 & 6.26 & 6.79 \\
 & 3$^1$A$_{g}$ ($\pi$ → $\pi^*$) & 6.21 & 7.46 & 6.90 & 6.90 \\
 & 3$^1$B$_{3u}$ ($\pi$ → $\pi^*$) & 7.68 & 9.05 & 8.41 & 8.12 \\
 & 3$^1$B$_{2u}$ ($\pi$ → $\pi^*$) & 7.55 & 8.24 & 8.04 & 8.44 \\[1.0ex]
Furan              & 1$^1$B$_{2}$ ($\pi$ → $\pi^*$) & 6.13 & 6.30 & 6.23 & 6.6 \\
 & 2$^1$A$_{1}$ ($\pi$ → $\pi^*$) & 6.39 & 6.96 & 6.33 & 6.62 \\
 & 3$^1$A$_{1}$ ($\pi$ → $\pi^*$) & 8.19 & 8.39 & 8.21 & 8.53 \\[1.0ex]
Pyrrole            & 2$^1$A$_{1}$ ($\pi$ → $\pi^*$) & 6.28 & 6.73 & 6.08 & 6.41 \\
 & 1$^1$B$_{2}$ ($\pi$ → $\pi^*$) & 6.37 & 6.52 & 6.38 & 6.71 \\
 & 3$^1$A$_{1}$ ($\pi$ → $\pi^*$) & 7.87 & 8.10 & 7.96 & 8.17 \\[1.0ex]
Imidazole          & 2$^1$A$'$ ($\pi$ → $\pi^*$) & 6.28 & 6.61 & 6.19 & 6.58 \\
 & 1$^1$A$''$ (n → $\pi^*$) & 5.88 & 6.62 & 6.32 & 6.83 \\
 & 3$^1$A$'$ ($\pi$ → $\pi^*$) & 6.38 & 7.20 & 6.73 & 7.10\\
 & 2$^1$A$''$ (n → $\pi^*$) & 7.06 & 7.64 & 7.54 & 7.94 \\
 & 4$^1$A$'$ ($\pi$ → $\pi^*$) & 8.05 & 8.43 & 8.12 & 8.45 \\[1.0ex]
Pyridine           & 1$^1$B$_{2}$ ($\pi$ → $\pi^*$) & 5.36 & 5.52 & 4.70 & 5.15 \\
 & 1$^1$B$_{1}$ (n → $\pi^*$) & 4.36 & 4.88 & 4.58 & 5.06 \\
 & 1$^1$A$_{2}$ (n → $\pi^*$) & 4.45 & 5.22 & 4.91 & 5.51 \\
 & 2$^1$A$_{1}$ ($\pi$ → $\pi^*$) & 6.23 & 6.51 & 6.31 & 6.85 \\
 & 3$^1$A$_{1}$ ($\pi$ → $\pi^*$) & 6.66 & 7.43 & 7.20 & 7.70 \\
 & 2$^1$B$_{2}$ ($\pi$ → $\pi^*$) & 7.14 & 7.37 & 7.09 & 7.59 \\
 & 3$^1$B$_{2}$ ($\pi$ → $\pi^*$) & 8.29 & 9.81 & 8.92 & 8.78 \\
 & 4$^1$A$_{1}$ ($\pi$ → $\pi^*$) & 7.28 & 8.32 & 8.76 & 8.68 \\[1.0ex]
Pyrazine           & 1$^1$B$_{3u}$ (n → $\pi^*$) & 3.56 & 4.03 & 3.72 & 4.25 \\
 & 1$^1$A$_{u}$ (n → $\pi^*$) & 4.03 & 4.79 & 4.50 & 5.05 \\
 & 1$^1$B$_{2u}$ ($\pi$ → $\pi^*$) & 5.26 & 5.36 & 4.48 & 5.02 \\
 & 1$^1$B$_{2g}$ (n → $\pi^*$) & 5.09 & 5.67 & 5.34 & 5.74 \\
 & 1$^1$B$_{1g}$ (n → $\pi^*$) & 5.54 & 6.58 & 6.24 & 6.76 \\
 & 1$^1$B$_{1u}$ ($\pi$ → $\pi^*$) & 6.44 & 6.69 & 6.52 & 7.07 \\
 & 2$^1$B$_{2u}$ ($\pi$ → $\pi^*$) & 7.36 & 7.88 & 7.53 & 8.05 \\
 & 2$^1$B$_{1u}$ ($\pi$ → $\pi^*$) & 7.52 & 7.81 & 7.53 & 8.06 \\
 & 1$^1$B$_{3g}$ ($\pi$ → $\pi^*$) & 8.17 & 10.03 & 8.93 & 8.77 \\
 & 2$^1$A$_{g}$ ($\pi$ → $\pi^*$) & 7.90 & 8.31 & 8.75 & 8.70 \\[1.0ex]
Pyrimidine         & 1$^1$B$_{1}$ (n → $\pi^*$) & 3.77 & 4.35 & 3.93 & 4.51 \\
 & 1$^1$A$_{2}$ (n → $\pi^*$) & 4.00 & 4.70 & 4.32 & 4.93 \\
 & 1$^1$B$_{2}$ ($\pi$ → $\pi^*$) & 5.59 & 5.75 & 4.83 & 5.37 \\
 & 2$^1$A$_{1}$ ($\pi$ → $\pi^*$) & 6.48 & 6.78 & 6.50 & 7.06 \\
 & 3$^1$A$_{1}$ ($\pi$ → $\pi^*$) & 7.33 & 7.56 & 7.17 & 7.74 \\
 & 2$^1$B$_{2}$ ($\pi$ → $\pi^*$) & 7.58 & 7.77 & 7.42 & 8.01 \\[1.0ex]
Pyridazine         & 1$^1$B$_{1}$ (n → $\pi^*$) & 3.12 & 3.63 & 3.31 & 3.93 \\
 & 1$^1$A$_{2}$ (n → $\pi^*$) & 3.51 & 4.28 & 3.91 & 4.50 \\
 & 2$^1$A$_{1}$ ($\pi$ → $\pi^*$) & 5.46 & 5.63 & 4.67 & 5.22 \\
 & 2$^1$A$_{2}$ (n → $\pi^*$) & 4.99 & 5.57 & 5.26 & 5.75 \\
 & 2$^1$B$_{1}$ (n → $\pi^*$) & 5.42 & 6.21 & 5.91 & 6.41 \\
 & 1$^1$B$_{2}$ ($\pi$ → $\pi^*$) & 5.96 & 6.64 & 6.38 & 6.93 \\
 & 2$^1$B$_{2}$ ($\pi$ → $\pi^*$) & 6.34 & 7.29 & 7.01 & 7.55 \\
 & 3$^1$A$_{1}$ ($\pi$ → $\pi^*$) & 5.46 & 7.60 & 7.35 & 7.82 \\[1.0ex]
s-Triazine           & 1$^1$A$''_1$  (n → $\pi^*$) & 3.81 & 4.53 & 4.12 & 4.78 \\
 & 1$^1$A$''_2$  (n → $\pi^*$) & 4.05 & 4.64 & 4.24 & 4.76 \\
 & 1$^1$A$'_2$  ($\pi$ → $\pi^*$) & 5.95 & 6.14 & 5.08 & 5.71 \\
 & 2$^1$A$'_1$  ($\pi$ → $\pi^*$) & 6.9 & 7.20 & 6.78 & 7.41 \\[1.0ex]
s-Tetrazine          & 1$^1$B$_{3u}$ (n → $\pi^*$) & 1.82 & 2.26 & 1.80 & 2.54 \\
 & 1$^1$A$_{u}$ ($\pi$ → $\pi^*$) & 2.82 & 3.61 & 3.19 & 3.80 \\
 & 1$^1$B$_{1g}$ (n → $\pi^*$) & 4.10 & 4.78 & 4.42 & 4.98 \\
 & 1$^1$B$_{2u}$ ($\pi$ → $\pi^*$) & 5.47 & 5.55 & 4.37 & 5.12 \\
 & 1$^1$B$_{2g}$ (n → $\pi^*$) & 4.77 & 5.41 & 4.89 & 5.34 \\
 & 2$^1$A$_{u}$ (n → $\pi^*$) & 4.58 & 5.16 & 4.88 & 5.46 \\
 & 2$^1$B$_{2g}$ (n → $\pi^*$) & 5.21 & 6.13 & 5.83 & 6.25 \\
 & 2$^1$B$_{1g}$ (n → $\pi^*$) & 5.84 & 6.79 & 6.41 & 6.87 \\
 & 3$^1$B$_{1g}$ (n → $\pi^*$) & 6.51 & 7.89 & 7.24 & 7.09 \\
 & 2$^1$B$_{3u}$ (n → $\pi^*$) & 5.61 & 6.39 & 6.21 & 6.68 \\
 & 1$^1$B$_{1u}$ ($\pi$ → $\pi^*$) & 6.84 & 7.12 & 6.84 & 7.45 \\
 & 2$^1$B$_{1u}$ ($\pi$ → $\pi^*$) & 7.37 & 7.52 & 7.15 & 7.79 \\
 & 2$^1$B$_{2u}$ ($\pi$ → $\pi^*$) & 8.10 & 8.33 & 7.98 & 8.51 \\
 & 2$^1$B$_{3g}$ ($\pi$ → $\pi^*$) & 8.73 & 9.11 & 8.35 & 8.48 \\[1.0ex]
Formaldehyde       & 1$^1$A$_{2}$ (n → $\pi^*$) & 3.77 & 3.77 & 3.45 & 3.95 \\
 & 1$^1$B$_{1}$ ($\sigma$ → $\pi^*$) & 8.78 & 8.96 & 8.70 & 9.19 \\
 & 3$^1$A$_{1}$ ($\sigma$ → $\pi^*$) & 9.94 & 10.39 & 9.55 & 10.45 \\[1.0ex]
Acetone            & 1$^1$A$_{2}$ (n → $\pi^*$) & 4.20 & 4.27 & 3.82 & 4.40 \\
 & 1$^1$B$_{1}$ ($\sigma$ → $\pi^*$) & 8.13 & 8.84 & 8.66 & 9.17 \\
 & 2$^1$A$_{1}$ ($\pi$ → $\pi^*$) & 8.49 & 9.59 & 8.96 & 9.65 \\[1.0ex]
p-Benzoquinone       & 1$^1$B$_{1g}$ (n → $\pi^*$) & 1.87 & 2.60 & 2.09 & 2.75 \\
 & 1$^1$A$_{u}$ (n → $\pi^*$) & 2.00 & 2.78 & 2.17 & 2.85 \\
 & 1$^1$B$_{3g}$ ($\pi$ → $\pi^*$) & 3.37 & 4.11 & 4.21 & 4.59 \\
 & 1$^1$B$_{1u}$ ($\pi$ → $\pi^*$) & 4.49 & 5.08 & 4.76 & 5.62 \\
 & 1$^1$B$_{3u}$ (n → $\pi^*$) & 4.36 & 5.91 & 5.22 & 5.83 \\
 & 2$^1$B$_{3g}$ ($\pi$ → $\pi^*$) & 6.12 & 7.09 & 6.74 & 7.28 \\
 & 2$^1$B$_{1u}$ ($\pi$ → $\pi^*$) & 6.83 & 7.74 & 7.74 & 7.82 \\[1.0ex]
Formamide          & 1$^1$A$''$ (n → $\pi^*$) & 5.45 & 5.45 & 5.00 & 5.66 \\
 & 2$^1$A$'$ ($\pi$ → $\pi^*$) & 6.15 & 7.51 & 7.48 & 7.23 \\[1.0ex]
Acetamide          & 1$^1$A$''$ (n → $\pi^*$) & 5.40 & 5.51 & 5.01 & 5.70 \\
 & 2$^1$A$'$ ($\pi$ → $\pi^*$) & 6.88 & 7.76 & 7.02 & 7.67 \\[1.0ex]
Propanamide        & 1$^1$A$''$ (n → $\pi^*$) & 5.42 & 5.53 & 5.02 & 5.72 \\
 & 2$^1$A$'$ ($\pi$ → $\pi^*$) & 6.54 & 7.69 & 6.96 & 7.62 \\[1.0ex]
Cytosine           & 2$^1$A$'$ ($\pi$ → $\pi^*$) & 4.20 & 4.83 & 4.06 & 4.72 \\
 & 1$^1$A$''$ (n → $\pi^*$) & 3.78 & 5.03 & 4.78 & 5.16 \\
 & 3$^1$A$'$ ($\pi$ → $\pi^*$) & 4.91 & 5.74 & 5.09 & 5.61 \\
 & 4$^1$A$'$ ($\pi$ → $\pi^*$) & 5.62 & 6.57 & 5.92 & 6.61 \\[1.0ex]
Thymine            & 1$^1$A$''$ (n → $\pi^*$) & 4.07 & 4.85 & 4.17 & 4.94 \\
 & 2$^1$A$'$ ($\pi$ → $\pi^*$) & 4.59 & 5.15 & 4.74 & 5.34 \\
 & 2$^1$A$''$ (n → $\pi^*$) & 4.78 & 6.19 & 5.62 & 6.59 \\
 & 4$^1$A$'$ ($\pi$ → $\pi^*$) & 5.83 & 6.54 & 6.24 & 6.71 \\[1.0ex]
Uracil             & 1$^1$A$''$ (n → $\pi^*$) & 3.95 & 4.83 & 4.15 & 4.90 \\
 & 2$^1$A$'$ ($\pi$ → $\pi^*$) & 4.76 & 5.34 & 4.87 & 5.44 \\
 & 2$^1$A$''$ (n → $\pi^*$) & 4.74 & 6.11 & 5.55 & 6.32 \\
 & 4$^1$A$'$ ($\pi$ → $\pi^*$) & 5.88 & 6.73 & 6.41 & 6.84 \\
 & 3$^1$A$''$ (n → $\pi^*$) & 5.22 & 6.80 & 6.16 & 6.87 \\
 & 5$^1$A$''$ (n → $\pi^*$) & 6.10 & 7.04 & 6.44 & 7.12 \\[1.0ex]
Adenine            & 2$^1$A$'$ ($\pi$ → $\pi^*$) & 4.57 & 5.23 & 4.62 & 5.18 \\
 & 1$^1$A$''$ (n → $\pi^*$) & 4.28 & 5.10 & 4.69 & 5.34 \\
 & 2$^1$A$''$ (n → $\pi^*$) & 5.03 & 5.73 & 5.32 & 5.96 \\
 & 4$^1$A$'$ ($\pi$ → $\pi^*$) & 5.56 & 6.49 & 5.99 & 6.53 \\

\end{longtable} \end{ThreePartTable} \normalsize
\begin{table}[ht!]
\centering 
\caption{Deviations of the vertical singlet excitation energies (in eV) in the TZVP basis set from the CC3/TZVP results.}
\label{tab:sing_tzvp_stat}
\small
\begin{tabular}{cccc} \toprule 
& PBE & srSOPPA & SOPPA \\
\midrule   &  & all &  \\
Count & 132 & 132 & 132 \\
Mean & -0.74 & -0.08 & -0.45 \\
Abs. mean & 0.77 & 0.30 & 0.50 \\
Std. dev. & 0.45 & 0.37 & 0.28 \\
Maximum (+) & 0.35 & 1.26 & 0.58 \\
Maximum (-) & 2.36 & 0.73 & 0.97 \\
\midrule   &  & ($\pi$ → $\pi^*$) &  \\
Count & 86 & 86 & 86 \\
Mean & -0.67 & -0.04 & -0.39 \\
Abs. mean & 0.72 & 0.35 & 0.46 \\
Std. dev. & 0.47 & 0.43 & 0.31 \\
Maximum (+) & 0.35 & 1.26 & 0.58 \\
Maximum (-) & 2.36 & 0.73 & 0.86 \\
\midrule   &  & (n → $\pi^*$) &  \\
Count & 42 & 42 & 42 \\
Mean & -0.89 & -0.16 & -0.57 \\
Abs. mean & 0.89 & 0.20 & 0.58 \\
Std. dev. & 0.36 & 0.18 & 0.16 \\
Maximum (+) & 0 & 0.80 & 0.15 \\
Maximum (-) & 1.81 & 0.40 & 0.97 \\

\bottomrule \end{tabular}\end{table} \normalsize


For SOPPA-srPBE, the largest outliers above the mean deviation from CC3 are the $2^1$A$_g$ state of all-E-hexatriene, $2^1$E$_{2g}$ state of benzene, $3^1$B$_{3u}$ state of naphtalene, $3^1$B$_2$ state of pyridine and the $1^1$B$_{3g}$ state of pyrazine. These states are all overestimated by more than $0.80$ eV compared to the CC3 results. They are also all states that have large double excitation character as seen by the singlet excitation character being less than 70\% \cite{noneiterative}. %For SOPPA these states are also outliers above the mean deviation from CC3, but SOPPA comes closer than SOPPA-srPBE for these states.
%PBE is behaving differently and the energies of these states are below the CC3 energies for PBE.

The largest outliers below the mean deviation from the CC3 for SOPPA-srPBE are the $1^1$B$_u$ states of E-butadiene, all-E-hexatriene and all-E-octatetraene and $1^1$B$_2$ state of cyclopropene. These states are all underestimated by more than $0.60$ eV compared to the CC3 results. SOPPA and PBE have far more states that underestimate the CC3 results by more than $0.60$ eV, which is expected from the more negative mean deviations.  The largest outliers for SOPPA are the $1^1$B$_u$ state of all-E-octatetraene, $3^1$A$_1$ state of formaldehyde, $1^1$B$_{1u}$ of p-benzoquinone and $2^1$A$''$ state of thymine. The largest outliers for PBE are the 3$^1$B$_{u}$ state of all-E-octatetraene, 3$^1$A$_{1}$ state of pyridazine, 2$^1$A$''$ state of thymine and 3$^1$A$''$ state of uracil. Overall the worst outliers below the mean deviation for SOPPA-srPBE are not the same as the worst outliers for SOPPA and PBE. 

The largest outliers are for SOPPA-srPBE primarily $\pi \rightarrow \pi^*$ transitions. The statistical analysis in table \ref{tab:sing_tzvp_stat} have therefore been evaluated separately for the $\pi \rightarrow \pi^*$ transitions and the $n \rightarrow \pi^*$ transitions. Compared to the whole dataset it can be seen that SOPPA-srPBE performs worse for the $\pi \rightarrow \pi^*$ transitions and better for the $n \rightarrow \pi^*$ transitions, where the standard deviation of SOPPA-srPBE comes close to the standard deviation of SOPPA. The other two methods have a smaller mean absolute deviation in for the $\pi \rightarrow \pi^*$ transitions but a smaller standard deviation for the $n \rightarrow \pi^*$ transitions. %still some tlc needed here with regards to the introductory sentence, return to later

SOPPA-srPBE outperformes SOPPA at 107 states out of 132 investigated states, and overall it performs the best. This can also be seen from table \ref{tab:sing_tzvp_stat} and figure \ref{fig:sing_corr}, as the srSOPPA method produced the lowest mean and absolute mean errors, and lies closest to the y axis. The biggest improvements are made for the  3$^1$A$_{1}$ state of formaldehyde, 1$^1$A$_{u}$ state of p-benzoquinone, 1$^1$A$^{''}$ states of thymine and uracil and 4$^1$A$^{'}$ state of cytosine.

Comparing to a popular alternative, B3LYP, for the same benchmarking set the mean deviation is $-0.08\pm0.29$ eV for SOPPA-srPBE and $-0.29\pm0.46$ eV for B3LYP \cite{dft2008}. Both the accuracy and consistency of SOPPA-srPBE is better than for B3LYP.

\subsection{Singlet excitation energies with the aug-cc-pVTZ basis set}

\begin{figure}[h!]
    \centering
    \subfloat{\includegraphics[width=5cm]{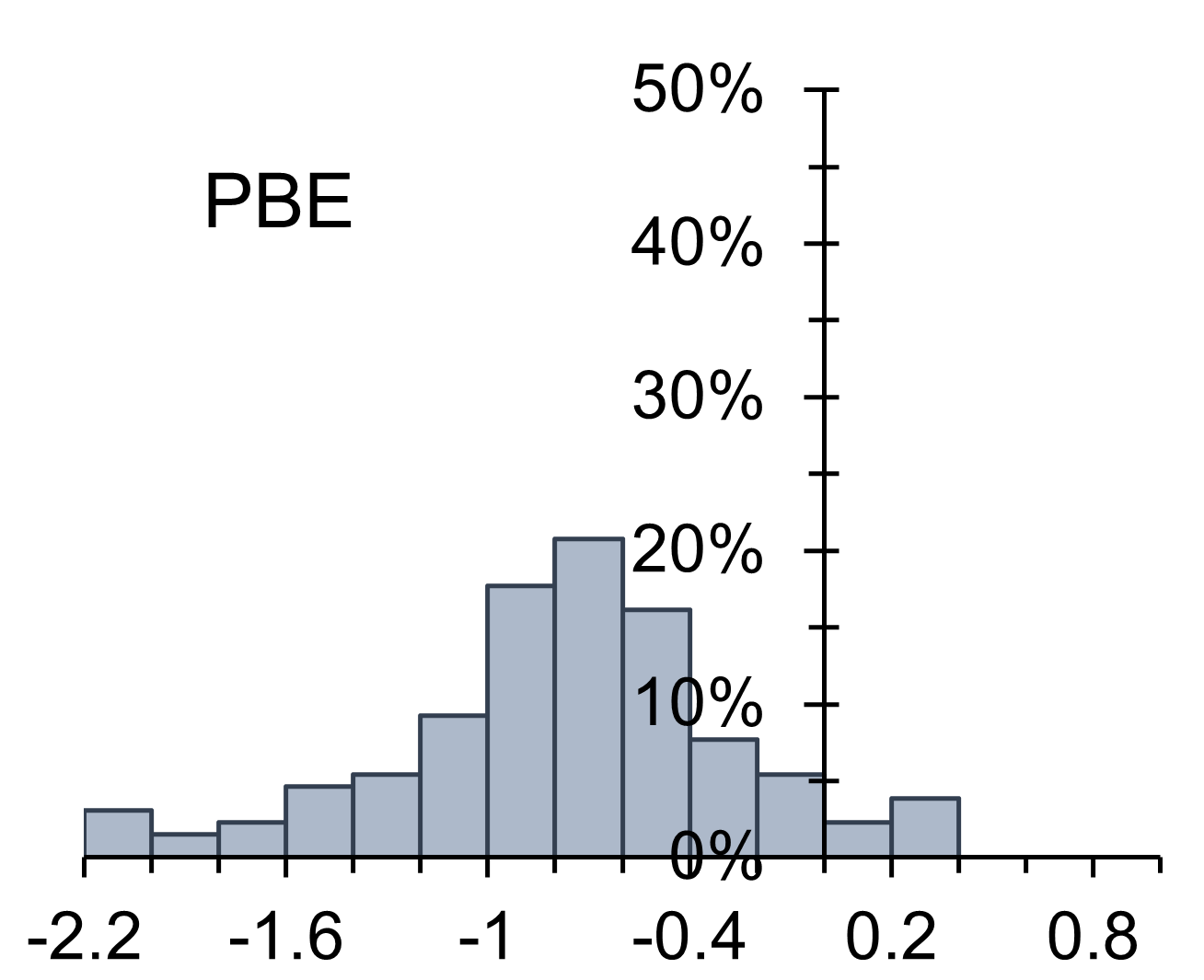}}
    \subfloat{\includegraphics[width=5cm]{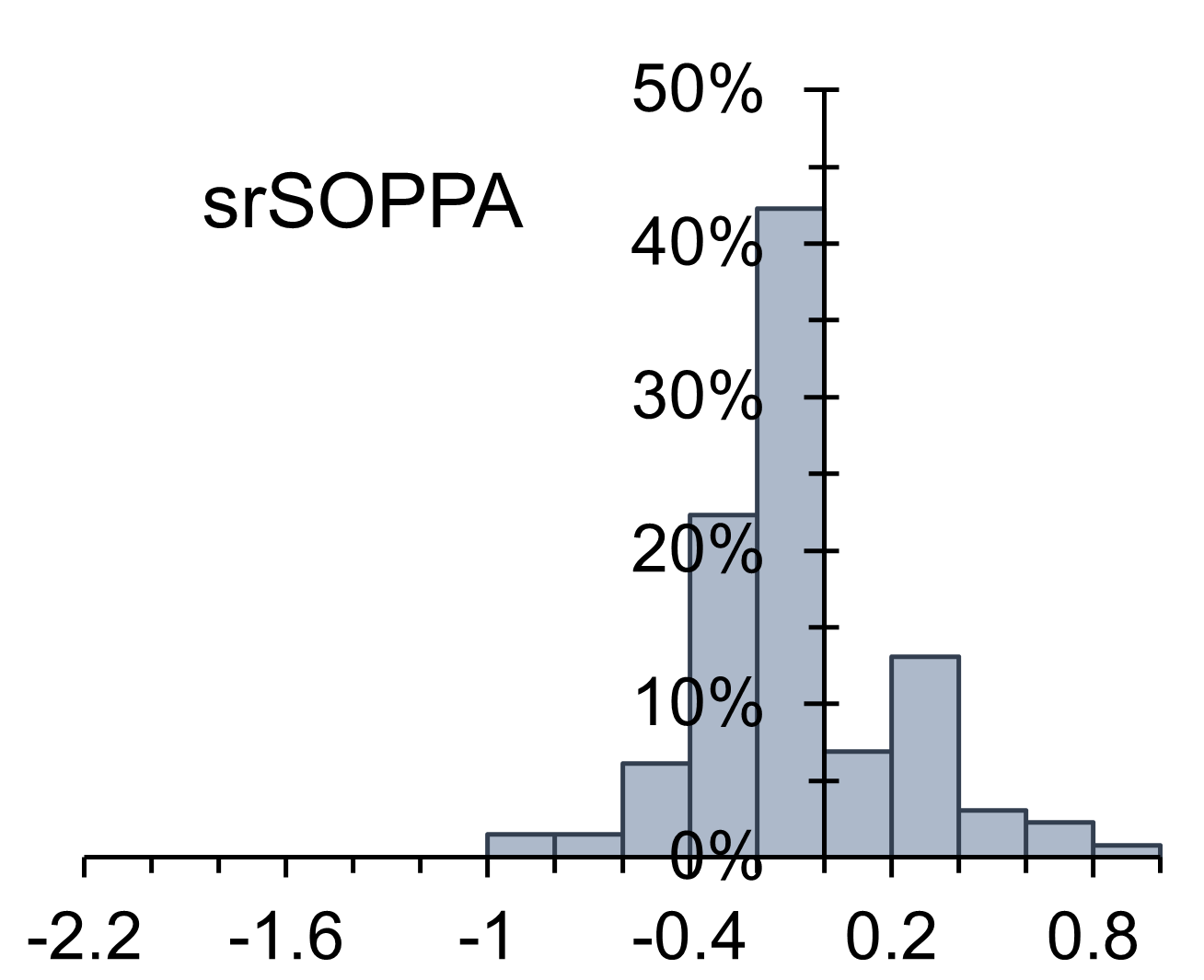}}
    \subfloat{\includegraphics[width=5cm]{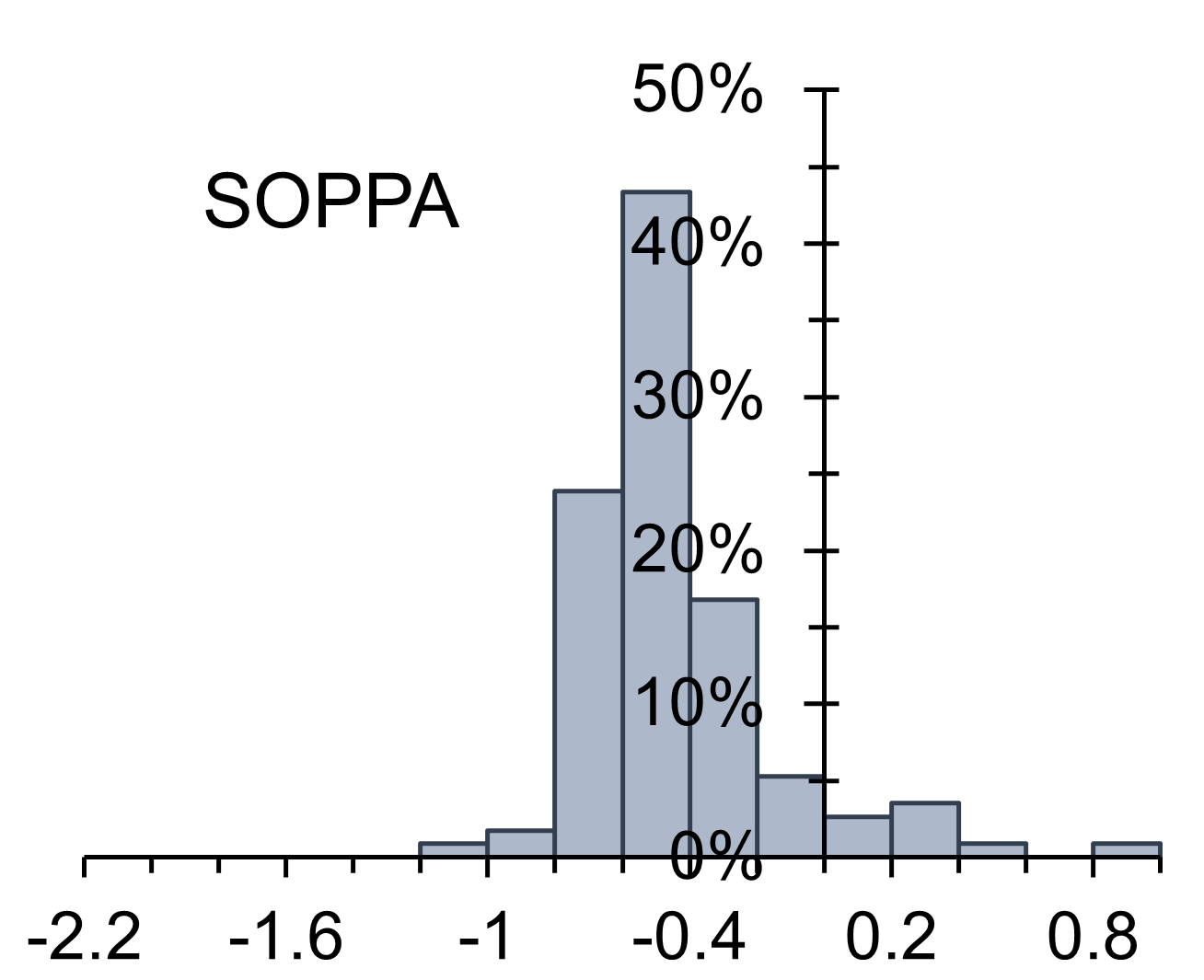}}
    \caption{Histograms (in \%) of the deviations of PBE, SOPPA-srPBE (srSOPPA) and SOPPA vertical singlet excitation energies (in eV) from the CC3 energies in the aug-cc-pVTZ basis set.}
    \label{fig:sing_dev_aug}
\end{figure}

\begin{figure}[h!]
    \centering
    \subfloat{\includegraphics[width=5cm]{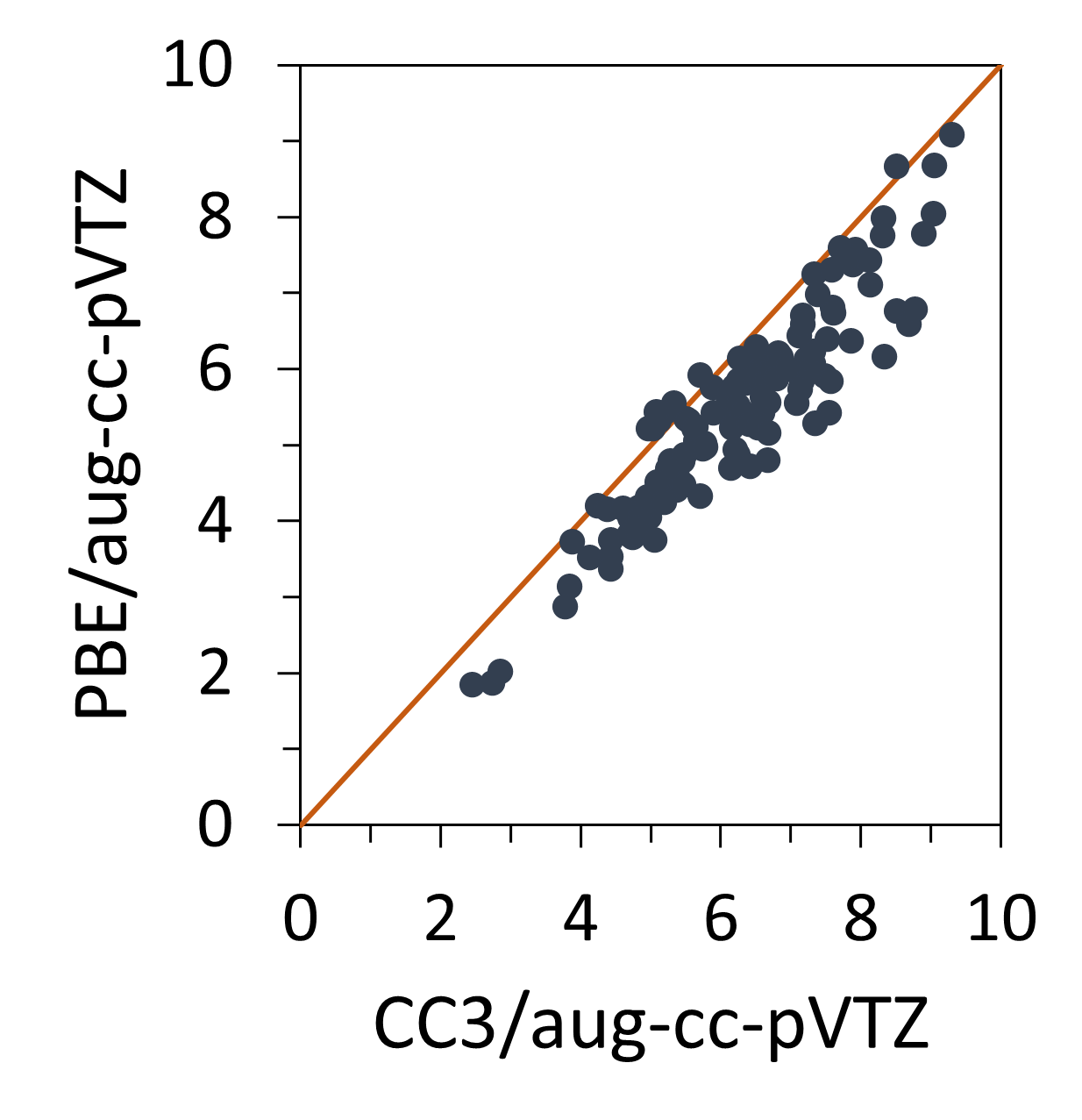}}
    \subfloat{\includegraphics[width=5cm]{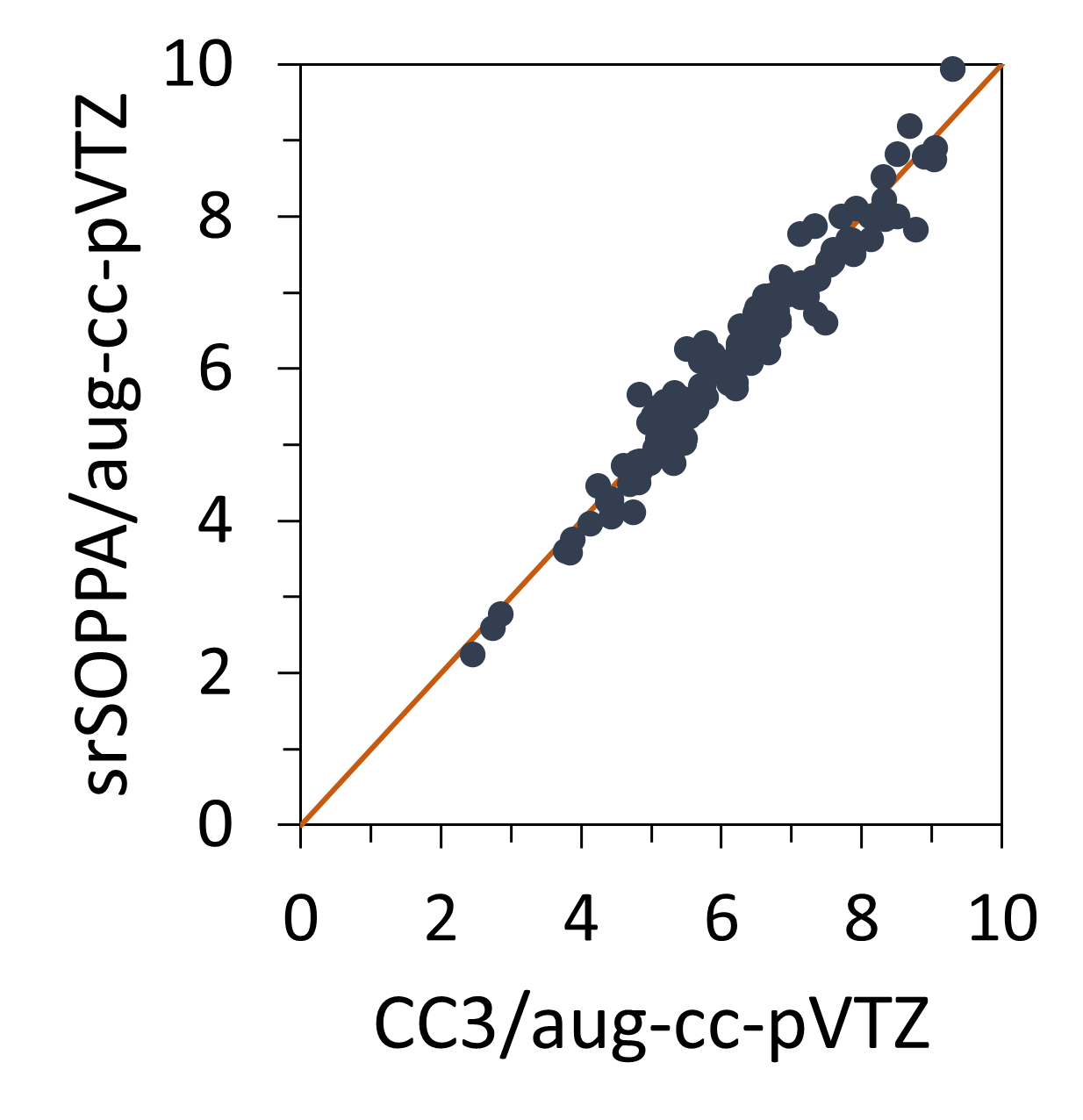}}
    \subfloat{\includegraphics[width=5cm]{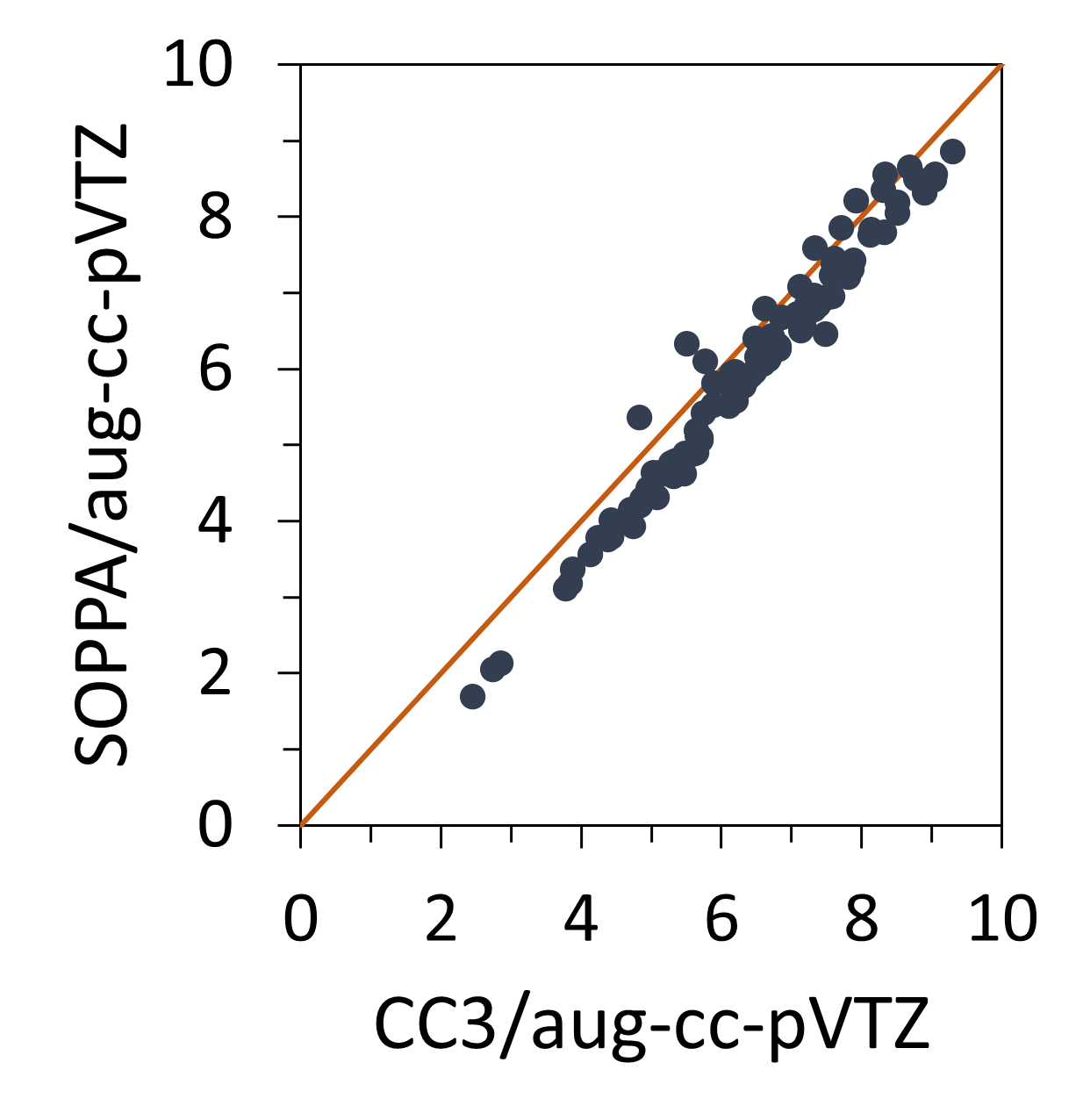}}
    \caption{Correlation plot of the vertical singlet excitation energies (in eV) with PBE, SOPPA-srPBE (srSOPPA) and SOPPA versus CC3 in the aug-cc-pVTZ basis set.}
    \label{fig:sing_corr_aug}
\end{figure}

The histograms of the deviations from the CC3 results are shown in figure \ref{fig:sing_dev_aug} for the PBE, SOPPA-srPBE and SOPPA results in the aug-cc-pVTZ basis set. The corresponding correlation plots are shown in figure \ref{fig:sing_corr_aug}. The statistical analysis in table \ref{tab:sing_avtz_stat} shows that the performance of the three methods in the aug-cc-pVTZ basis set is similar to the performance in the TZVP basis set. The PBE and the SOPPA methods still underestimate the excitation energies compared to the CC3 results, apart from a few outliers. SOPPA-srPBE both under- and overestimates the energies as before. SOPPA-srPBE has the smallest absolute mean deviation ($0.24$ eV), followed by SOPPA ($0.50$ eV) and lastly PBE ($0.81$ eV). For the standard deviations SOPPA is the smallest ($0.28$ eV), followed by SOPPA-srPBE ($0.29$ eV) and then PBE ($0.52$ eV). This is the same ranking as for the TZVP basis set, although the standard deviation of srSOPPA-PBE is basically the same with the standard deviation of SOPPA this time. Overall for the singlet states SOPPA-srPBE performs the best.
\small
\begin{ThreePartTable}
\begin{TableNotes}
    \item[a] Results from \cite{soppa}
    \item[b] Results from \cite{cc32010} if not otherwise marked
    \item[c] CC3/TZVP result from \cite{cc32008} with basis set correction from CCSDR(3)/aug-cc-pVTZ
and CCSDR(3)/TZVP from \cite{cc32010}
    \item[d] CC3/TZVP result from \cite{cc32008} or \cite{cc32014} with basis set correction from CC2/aug-cc-pVTZ and CC2/TZVP from \cite{cc32010}
\end{TableNotes}
\begin{longtable}[h!]{p{2.7cm}p{2.6cm}cccc}
\caption{Vertical singlet excitation energies (in eV) with the aug-cc-pVTZ basis set.}
\label{tab:sing_aVTZ}

\\ \toprule Molecule & State & PBE & srSOPPA & SOPPA\tnote{a} & CC3\tnote{b} \\ \midrule \endfirsthead

\multicolumn{6}{l}{{\tablename} \thetable{}: \textit{continued.}} \\
\toprule Molecule & State & PBE & srSOPPA & SOPPA\tnote{a} & CC3\tnote{b} \\ \midrule\endhead

\bottomrule \multicolumn{6}{r}{\textit{continued.}} \endfoot 

\bottomrule \insertTableNotes \endlastfoot
Ethene & 1$^1$B$_{1u}$ ($\pi$ → $\pi^*$) & 7.37 & 7.50 & 7.42 & 7.89 \\[1.0ex]
E-Butadiene & 1$^1$B$_{u}$ ($\pi$ → $\pi^*$) & 5.43 & 5.73 & 5.58 & 6.21\tnote{c} \\
 & 2$^1$A$_{g}$ ($\pi$ → $\pi^*$) & 6.10 & 6.94 & 6.79 & 6.63 \\[1.0ex]
all-E-Hexatriene & 1$^1$B$_{u}$ ($\pi$ → $\pi^*$) & 4.43 & 4.75 & 4.58 & 5.32\tnote{c} \\
 & 2$^1$A$_{g}$ ($\pi$ → $\pi^*$) & 5.02 & 6.33 & 6.09 & 5.77\tnote{c} \\[1.0ex]
all-E-Octatetraene & 2$^1$A$_{g}$ ($\pi$ → $\pi^*$) & 4.17 & 5.65 & 5.36 & 4.84\tnote{d} \\
 & 1$^1$B$_{u}$ ($\pi$ → $\pi^*$) & 3.78 & 4.11 & 3.92 & 4.75\tnote{d} \\
 & 2$^1$B$_{u}$ ($\pi$ → $\pi^*$) & 5.34 & 6.25 & 6.33 & 5.51\tnote{d} \\
 & 3$^1$A$_{g}$ ($\pi$ → $\pi^*$) & 5.42 & 5.95 & 5.81 & 5.90\tnote{d} \\
 & 4$^1$A$_{g}$ ($\pi$ → $\pi^*$) & 5.69 & 6.06 & 5.93 & 6.15\tnote{d} \\
 & 3$^1$B$_{u}$ ($\pi$ → $\pi^*$) & 5.81 & 6.46 & 6.97 &  \\[1.0ex]
Cyclopropene      & 1$^1$B$_{1}$ ($\sigma$ → $\pi^*$) & 5.87 & 6.39 & 6.34 & 6.67 \\
 & 1$^1$B$_{2}$ ($\pi$ → $\pi^*$) & 5.90 & 6.21 & 6.24 & 6.68 \\[1.0ex]
Cyclopentadiene    & 1$^1$B$_{2}$ ($\pi$ → $\pi^*$) & 4.86 & 5.08 & 4.88 & 5.49\tnote{c} \\
 & 2$^1$A$_{1}$ ($\pi$ → $\pi^*$) & 6.00 & 6.73 & 6.39 & 6.49\tnote{c} \\
 & 3$^1$A$_{1}$ ($\pi$ → $\pi^*$) & 7.10 & 7.69 & 7.82 & 8.14\tnote{c} \\[1.0ex]
Norbornadiene      & 1$^1$A$_{2}$ ($\pi$ → $\pi^*$) & 4.40 & 4.98 & 4.79 & 5.37\tnote{d} \\
 & 1$^1$B$_{2}$ ($\pi$ → $\pi^*$) & 4.93 & 5.81 & 5.68 & 6.21\tnote{d} \\
 & 2$^1$B$_{2}$ ($\pi$ → $\pi^*$) & 5.90 & 6.60 & 6.45 & 7.49\tnote{d} \\
 & 2$^1$A$_{2}$ ($\pi$ → $\pi^*$) & 6.12 & 6.95 & 6.84 & 7.22\tnote{d} \\[1.0ex]
Benzene            & 1$^1$B$_{2u}$ ($\pi$ → $\pi^*$) & 5.21 & 5.38 & 4.63 & 5.03 \\
 & 1$^1$B$_{1u}$ ($\pi$ → $\pi^*$) & 5.94 & 6.17 & 5.91 & 6.42 \\
 & 1$^1$E$_{1u}$ ($\pi$ → $\pi^*$) & 5.94 & 6.93 & 6.67 & 7.14 \\
 & 2$^1$E$_{2g}$ ($\pi$ → $\pi^*$) & 7.75 & 8.52 & 8.34 & 8.31 \\[1.0ex]
Naphthalene        & 1$^1$B$_{3u}$ ($\pi$ → $\pi^*$) & 4.20 & 4.45 & 3.78 & 4.25\tnote{c} \\
 & 1$^1$B$_{2u}$ ($\pi$ → $\pi^*$) & 4.03 & 4.50 & 4.19 & 4.82\tnote{c} \\
 & 2$^1$A$_{g}$ ($\pi$ → $\pi^*$) & 5.75 & 6.18 & 5.52 & 5.89\tnote{c} \\
 & 1$^1$B$_{1g}$ ($\pi$ → $\pi^*$) & 4.95 & 5.75 & 5.41 & 5.75\tnote{c} \\
 & 2$^1$B$_{3u}$ ($\pi$ → $\pi^*$) & 5.63 & 5.80 & 5.50 & 6.11\tnote{c} \\
 & 2$^1$B$_{2u}$ ($\pi$ → $\pi^*$) & 5.81 & 6.19 & 5.85 & 6.36\tnote{c} \\
 & 2$^1$B$_{1g}$ ($\pi$ → $\pi^*$) & 5.89 & 6.32 & 5.96 & 6.47\tnote{c} \\
 & 3$^1$A$_{g}$ ($\pi$ → $\pi^*$) & 6.16 & 7.20 & 6.67 & 6.86\tnote{c} \\
 & 3$^1$B$_{3u}$ ($\pi$ → $\pi^*$) & 7.56 & 8.10 & 8.21 & 7.93\tnote{c} \\
 & 3$^1$B$_{2u}$ ($\pi$ → $\pi^*$) & 7.24 & 7.87 & 7.58 & 7.34\tnote{c} \\[1.0ex]
Furan              & 1$^1$B$_{2}$ ($\pi$ → $\pi^*$) & 5.85 & 6.06 & 5.89 & 6.26\tnote{c} \\
 & 2$^1$A$_{1}$ ($\pi$ → $\pi^*$) & 6.28 & 6.80 & 6.15 & 6.51\tnote{c} \\
 & 3$^1$A$_{1}$ ($\pi$ → $\pi^*$) & 7.43 & 7.99 & 7.76 & 8.13\tnote{c} \\[1.0ex]
Pyrrole            & 2$^1$A$_{1}$ ($\pi$ → $\pi^*$) & 6.13 & 6.55 & 5.90 & 6.27\tnote{c} \\
 & 1$^1$B$_{2}$ ($\pi$ → $\pi^*$) & 5.77 & 6.12 & 5.95 & 6.20\tnote{c} \\
 & 3$^1$A$_{1}$ ($\pi$ → $\pi^*$) & 6.8 & 7.56 & 7.37 & 7.60\tnote{c} \\[1.0ex]
Imidazole          & 2$^1$A$'$ ($\pi$ → $\pi^*$) & 5.51 & 6.32 & 5.92 & 6.25\tnote{d} \\
 & 1$^1$A$''$ (n → $\pi^*$) & 5.84 & 6.52 & 6.14 & 6.65\tnote{d} \\
 & 3$^1$A$'$ ($\pi$ → $\pi^*$) & 5.99 & 6.96 & 6.43 & 6.73\tnote{d} \\
 & 2$^1$A$''$ (n → $\pi^*$) & 5.84 & 7.46 & 7.23 & 7.58\tnote{d} \\
 & 4$^1$A$'$ ($\pi$ → $\pi^*$) & 6.76 & 8 & 8.05 & 8.51\tnote{d} \\[1.0ex]
Pyridine           & 1$^1$B$_{2}$ ($\pi$ → $\pi^*$) & 5.31 & 5.45 & 4.63 & 5.12\tnote{c} \\
 & 1$^1$B$_{1}$ (n → $\pi^*$) & 4.31 & 4.80 & 4.42 & 4.96\tnote{c} \\
 & 1$^1$A$_{2}$ (n → $\pi^*$) & 4.45 & 5.16 & 4.77 & 5.41\tnote{c} \\
 & 2$^1$A$_{1}$ ($\pi$ → $\pi^*$) & 5.42 & 6.37 & 6.06 & 6.60\tnote{c} \\
 & 3$^1$A$_{1}$ ($\pi$ → $\pi^*$) & 6.07 & 7.19 & 6.96 & 7.33\tnote{c} \\
 & 2$^1$B$_{2}$ ($\pi$ → $\pi^*$) & 6.98 & 7.18 & 6.83 & 7.39\tnote{c} \\
 & 3$^1$B$_{2}$ ($\pi$ → $\pi^*$) & 7.59 & 8.00 & 7.85 & 7.72\tnote{c} \\
 & 4$^1$A$_{1}$ ($\pi$ → $\pi^*$) & 6.16 & 7.96 & 8.55 & 8.34\tnote{c} \\[1.0ex]
Pyrazine           & 1$^1$B$_{3u}$ (n → $\pi^*$) & 3.52 & 3.96 & 3.55 & 4.13\tnote{c} \\
 & 1$^1$A$_{u}$ (n → $\pi^*$) & 4.05 & 4.75 & 4.40 & 4.98\tnote{c} \\
 & 1$^1$B$_{2u}$ ($\pi$ → $\pi^*$) & 5.21 & 5.28 & 4.40 & 4.97\tnote{c} \\
 & 1$^1$B$_{2g}$ (n → $\pi^*$) & 5.05 & 5.60 & 5.18 & 5.65\tnote{c} \\
 & 1$^1$B$_{1g}$ (n → $\pi^*$) & 5.56 & 6.52 & 6.14 & 6.69\tnote{c} \\
 & 1$^1$B$_{1u}$ ($\pi$ → $\pi^*$) & 6.20 & 6.56 & 6.29 & 6.83\tnote{c} \\
 & 2$^1$B$_{2u}$ ($\pi$ → $\pi^*$) & 7.44 & 7.70 & 7.32 & 7.81\tnote{c} \\
 & 2$^1$B$_{1u}$ ($\pi$ → $\pi^*$) & 6.36 & 7.68 & 7.31 & 7.86\tnote{c} \\
 & 1$^1$B$_{3g}$ ($\pi$ → $\pi^*$) & 6.59 & 9.18 & 8.65 & 8.69\tnote{c} \\
 & 2$^1$A$_{g}$ ($\pi$ → $\pi^*$) & 6.78 & 7.82 & 8.50 & 8.78\tnote{c} \\[1.0ex]
Pyrimidine         & 1$^1$B$_{1}$ (n → $\pi^*$) & 3.75 & 4.28 & 3.79 & 4.43\tnote{c} \\
 & 1$^1$A$_{2}$ (n → $\pi^*$) & 4.00 & 4.65 & 4.20 & 4.85\tnote{c} \\
 & 1$^1$B$_{2}$ ($\pi$ → $\pi^*$) & 5.55 & 5.68 & 4.78 & 5.34\tnote{c} \\
 & 2$^1$A$_{1}$ ($\pi$ → $\pi^*$) & 6.07 & 6.63 & 6.25 & 6.82\tnote{c} \\
 & 3$^1$A$_{1}$ ($\pi$ → $\pi^*$) & 6.39 & 7.40 & 6.94 & 7.53\tnote{c} \\
 & 2$^1$B$_{2}$ ($\pi$ → $\pi^*$) & 7.46 & 7.63 & 7.20 & 7.81\tnote{c} \\[1.0ex]
Pyridazine         & 1$^1$B$_{1}$ (n → $\pi^*$) & 3.13 & 3.58 & 3.17 & 3.85\tnote{c} \\
 & 1$^1$A$_{2}$ (n → $\pi^*$) & 3.53 & 4.24 & 3.79 & 4.44\tnote{c} \\
 & 2$^1$A$_{1}$ ($\pi$ → $\pi^*$) & 5.42 & 5.56 & 4.62 & 5.20\tnote{c} \\
 & 2$^1$A$_{2}$ (n → $\pi^*$) & 4.98 & 5.52 & 5.11 & 5.66\tnote{c} \\
 & 2$^1$B$_{1}$ (n → $\pi^*$) & 5.42 & 6.14 & 5.77 & 6.33\tnote{c} \\
 & 1$^1$B$_{2}$ ($\pi$ → $\pi^*$) & 4.80 & 6.49 & 6.12 & 6.67\tnote{c} \\
 & 2$^1$B$_{2}$ ($\pi$ → $\pi^*$) & 6.23 & 7.15 & 6.78 & 7.33\tnote{c} \\
 & 3$^1$A$_{1}$ ($\pi$ → $\pi^*$) & 5.42 & 7.36 & 6.96 & 7.55\tnote{c} \\[1.0ex]
s-Triazine           & 1$^1$A$''_1$  (n → $\pi^*$) & 3.82 & 4.48 & 4.02 & 4.70\tnote{c} \\
 & 1$^1$A$''_2$  (n → $\pi^*$) & 4.04 & 4.61 & 4.14 & 4.71\tnote{c} \\
 & 1$^1$A$'_2$  ($\pi$ → $\pi^*$) & 5.92 & 6.09 & 5.05 & 5.71\tnote{c} \\
 & 2$^1$A$'_1$  ($\pi$ → $\pi^*$) & 6.70 & 7.05 & 6.55 & 7.18\tnote{c} \\[1.0ex]
s-Tetrazine          & 1$^1$B$_{3u}$ (n → $\pi^*$) & 1.84 & 2.24 & 1.69 & 2.46 \\
 & 1$^1$A$_{u}$ ($\pi$ → $\pi^*$) & 2.87 & 3.60 & 3.11 & 3.78\tnote{c} \\
 & 1$^1$B$_{1g}$ (n → $\pi^*$) & 4.11 & 4.74 & 4.28 & 4.87 \\
 & 1$^1$B$_{2u}$ ($\pi$ → $\pi^*$) & 5.43 & 5.48 & 4.31 & 5.08 \\
 & 1$^1$B$_{2g}$ (n → $\pi^*$) & 4.78 & 5.38 & 4.76 & 5.28 \\
 & 2$^1$A$_{u}$ (n → $\pi^*$) & 4.58 & 5.13 & 4.75 & 5.39\tnote{c} \\
 & 2$^1$B$_{2g}$ (n → $\pi^*$) & 5.22 & 6.10 & 5.72 & 6.16 \\
 & 2$^1$B$_{1g}$ (n → $\pi^*$) & 5.87 & 6.75 & 6.31 & 6.80 \\
 & 3$^1$B$_{1g}$ (n → $\pi^*$) & 6.44 & 7.76 & 7.07 & 7.12\tnote{c} \\
 & 2$^1$B$_{3u}$ (n → $\pi^*$) & 5.63 & 6.35 & 6.10 & 6.60 \\
 & 1$^1$B$_{1u}$ ($\pi$ → $\pi^*$) & 6.59 & 6.96 & 6.59 & 7.18 \\
 & 2$^1$B$_{1u}$ ($\pi$ → $\pi^*$) & 7.30 & 7.39 & 6.95 & 7.59 \\
 & 2$^1$B$_{2u}$ ($\pi$ → $\pi^*$) & 7.98 & 8.21 & 7.79 & 8.33\tnote{c} \\
 & 2$^1$B$_{3g}$ ($\pi$ → $\pi^*$) & 8.66 & 8.82 & 8.18 & 8.51\tnote{c} \\[1.0ex]
Formaldehyde       & 1$^1$A$_{2}$ (n → $\pi^*$) & 3.72 & 3.75 & 3.36 & 3.88 \\
 & 1$^1$B$_{1}$ ($\sigma$ → $\pi^*$) & 8.67 & 8.89 & 8.55 & 9.05 \\
 & 3$^1$A$_{1}$ ($\sigma$ → $\pi^*$) & 9.08 & 9.93 & 8.85 & 9.31 \\[1.0ex]
Acetone            & 1$^1$A$_{2}$ (n → $\pi^*$) & 4.15 & 4.25 & 3.75 & 4.38 \\
 & 1$^1$B$_{1}$ ($\sigma$ → $\pi^*$) & 8.04 & 8.74 & 8.48 & 9.04\tnote{c} \\
 & 2$^1$A$_{1}$ ($\pi$ → $\pi^*$) & 7.77 & 8.78 & 8.31 & 8.90 \\[1.0ex]
p-Benzoquinone       & 1$^1$B$_{1g}$ (n → $\pi^*$) & 1.87 & 2.58 & 2.04 & 2.74\tnote{c} \\
 & 1$^1$A$_{u}$ (n → $\pi^*$) & 2.01 & 2.77 & 2.12 & 2.86\tnote{c} \\
 & 1$^1$B$_{3g}$ ($\pi$ → $\pi^*$) & 3.36 & 4.05 & 4.01 & 4.44\tnote{c} \\
 & 1$^1$B$_{1u}$ ($\pi$ → $\pi^*$) & 4.47 & 5.02 & 4.62 & 5.47\tnote{c} \\
 & 1$^1$B$_{3u}$ (n → $\pi^*$) & 4.32 & 5.77 & 5.09 & 5.71\tnote{c} \\
 & 2$^1$B$_{3g}$ ($\pi$ → $\pi^*$) & 5.84 & 6.94 & 6.58 & 7.16\tnote{c} \\
 & 2$^1$B$_{1u}$ ($\pi$ → $\pi^*$) & 6.74 & 7.54 & 7.45 & 7.62\tnote{c} \\[1.0ex]
Formamide          & 1$^1$A$''$ (n → $\pi^*$) & 5.31 & 5.36 & 4.86 & 5.55 \\
 & 2$^1$A$'$ ($\pi$ → $\pi^*$) & 5.28 & 6.71 & 6.84 & 7.35 \\[1.0ex]
Acetamide          & 1$^1$A$''$ (n → $\pi^*$) & 5.21 & 5.42 & 4.88 & 5.62\tnote{c} \\
 & 2$^1$A$'$ ($\pi$ → $\pi^*$) & 5.73 & 7.12 & 6.50 & 7.14\tnote{c} \\[1.0ex]
Propanamide        & 1$^1$A$''$ (n → $\pi^*$) & 5.23 & 5.44 & 4.89 & 5.65\tnote{c} \\
 & 2$^1$A$'$ ($\pi$ → $\pi^*$) & 5.55 & 7.05 & 6.72 & 7.09\tnote{c} \\[1.0ex]
Cytosine           & 2$^1$A$'$ ($\pi$ → $\pi^*$) & 4.16 & 4.72 &  & 4.61\tnote{d} \\
 & 1$^1$A$''$ (n → $\pi^*$) & 3.75 & 4.95 &  & 5.06\tnote{d} \\
 & 3$^1$A$'$ ($\pi$ → $\pi^*$) & 4.79 & 5.60 &  & 5.46\tnote{d} \\
 & 4$^1$A$'$ ($\pi$ → $\pi^*$) & 4.87 & 6.27 &  & 6.25\tnote{d} \\[1.0ex]
Thymine            & 1$^1$A$''$ (n → $\pi^*$) & 4.01 & 4.77 &  & 4.83\tnote{d} \\
 & 2$^1$A$'$ ($\pi$ → $\pi^*$) & 4.50 & 4.99 &  & 5.14\tnote{d} \\
 & 2$^1$A$''$ (n → $\pi^*$) & 4.72 & 6.07 &  & 6.42\tnote{d} \\
 & 4$^1$A$'$ ($\pi$ → $\pi^*$) & 5.27 & 6.32 &  & 6.41\tnote{d} \\[1.0ex]
Uracil             & 1$^1$A$''$ (n → $\pi^*$) & 3.91 & 4.76 &  & 4.80\tnote{d} \\
 & 2$^1$A$'$ ($\pi$ → $\pi^*$) & 4.68 & 5.18 &  & 5.25\tnote{d} \\
 & 2$^1$A$''$ (n → $\pi^*$) & 4.69 & 6.00 &  & 6.15\tnote{d} \\
 & 4$^1$A$'$ ($\pi$ → $\pi^*$) & 5.22 & 6.50 &  & 6.54\tnote{d} \\
 & 3$^1$A$''$ (n → $\pi^*$) & 5.15 & 6.65 &  & 6.69\tnote{d} \\
 & 5$^1$A$''$ (n → $\pi^*$) & 6.00 & 6.97 &  & 6.95\tnote{d} \\[1.0ex]
Adenine            & 2$^1$A$'$ ($\pi$ → $\pi^*$) & 4.51 & 5.08 &  & 5.1\tnote{d} \\
 & 1$^1$A$''$ (n → $\pi^*$) & 4.24 & 5.02 &  & 5.2\tnote{d} \\
 & 2$^1$A$''$ (n → $\pi^*$) & 4.97 & 5.62 &  & 5.79\tnote{d} \\
 & 4$^1$A$'$ ($\pi$ → $\pi^*$) & 5.14 & 6.27 &  &  \\

\end{longtable} \end{ThreePartTable} \normalsize
\begin{table}[h!]
\centering 
\caption{Deviations of the vertical singlet excitation energies (in eV) in the aug-cc-pVTZ basis set from the CC3/aug-cc-pVTZ results.}
\label{tab:sing_avtz_stat}
\small
\begin{tabular}{cccc} \toprule 
& PBE & srSOPPA & SOPPA \\
\midrule   &  & all &  \\
Count & 130 & 130 & 113 \\
Mean & -0.78 & -0.08 & -0.45 \\
Abs. mean & 0.81 & 0.24 & 0.50 \\
Std. dev. & 0.52 & 0.29 & 0.28 \\
Maximum (+) & 0.35 & 0.81 & 0.82 \\
Maximum (-) & 2.18 & 0.96 & 1.04 \\
\midrule   &  & ($\pi$ → $\pi^*$) &  \\
Count & 84 & 84 & 76 \\
Mean & -0.75 & -0.06 & -0.39 \\
Abs. mean & 0.79 & 0.28 & 0.46 \\
Std. dev. & 0.59 & 0.34 & 0.31 \\
Maximum (+) & 0.35 & 0.81 & 0.82 \\
Maximum (-) & 2.18 & 0.96 & 1.04 \\
\midrule   &  & (n → $\pi^*$) &  \\
Count & 42 & 42 & 33 \\
Mean & -0.85 & -0.13 & -0.58 \\
Abs. mean & 0.85 & 0.17 & 0.58 \\
Std. dev. & 0.35 & 0.15 & 0.13 \\
Maximum (+) & 0 & 0.64 & 0 \\
Maximum (-) & 1.74 & 0.35 & 0.77 \\

\bottomrule \end{tabular}\end{table} \normalsize


The largest outliers for SOPPA-srPBE are the $2^1$B$_2$ state of norbornadine, $2^1$A$_g$ state of all-E-octatetraene and the $2^1$A$_g$ state of pyrazine. These states all have deviations larger than $0.79$ eV. For PBE the biggest outliers are the  $4^1$A$_1$ state of Pyridine,  $1^1$B$_{3g}$ and $2^1$A$_g$ state of pyrazine, $3^1$A$_1$ state of pyridazine, and  $2^1$A$'$ state of formamide, where the absolute deviation is more than $2.00$ eV. For SOPPA the largest outliers are the $1^1$B$_u$ and $2^1$B$_u$ state of all-E-octatetraene, $2^1$B$_2$ state of norbornadine, and  $1^1$B$_u$ state of p-benzoquinone, where the absolute deviation is more then $0.80$ eV for all of them. For all three methods the largest outliers have changed with the addition of the diffuse functions.

The largest contribution for the mean, absolute mean, and standard deviations for the srSOPPA-PBE method comes from the $\pi \rightarrow \pi^*$ transitions like before.

\subsection{Triplet excitation energies with the TZVP basis set}
\begin{figure}[h!]
    \centering
    \subfloat{\includegraphics[width=5cm]{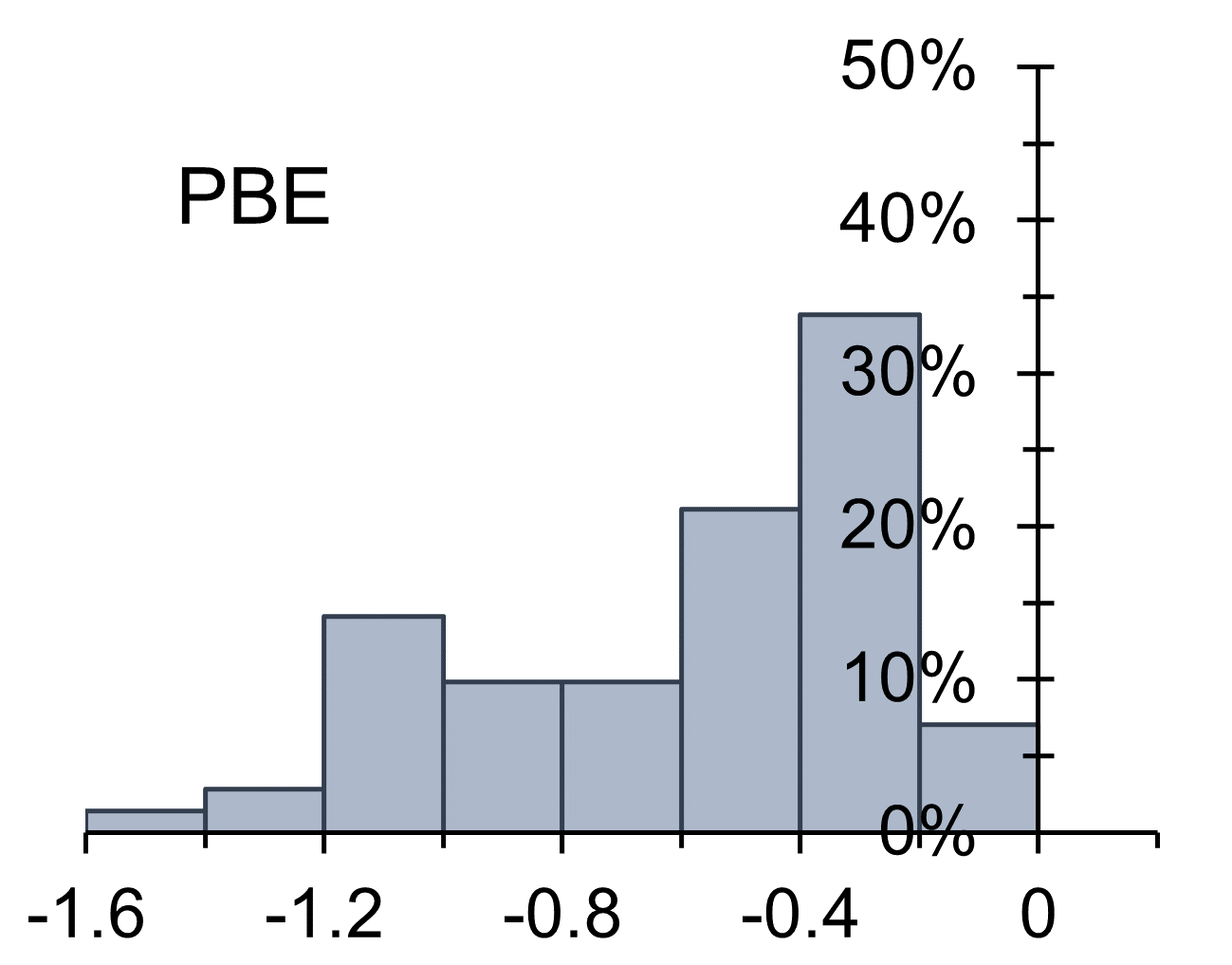}}
    \subfloat{\includegraphics[width=5cm]{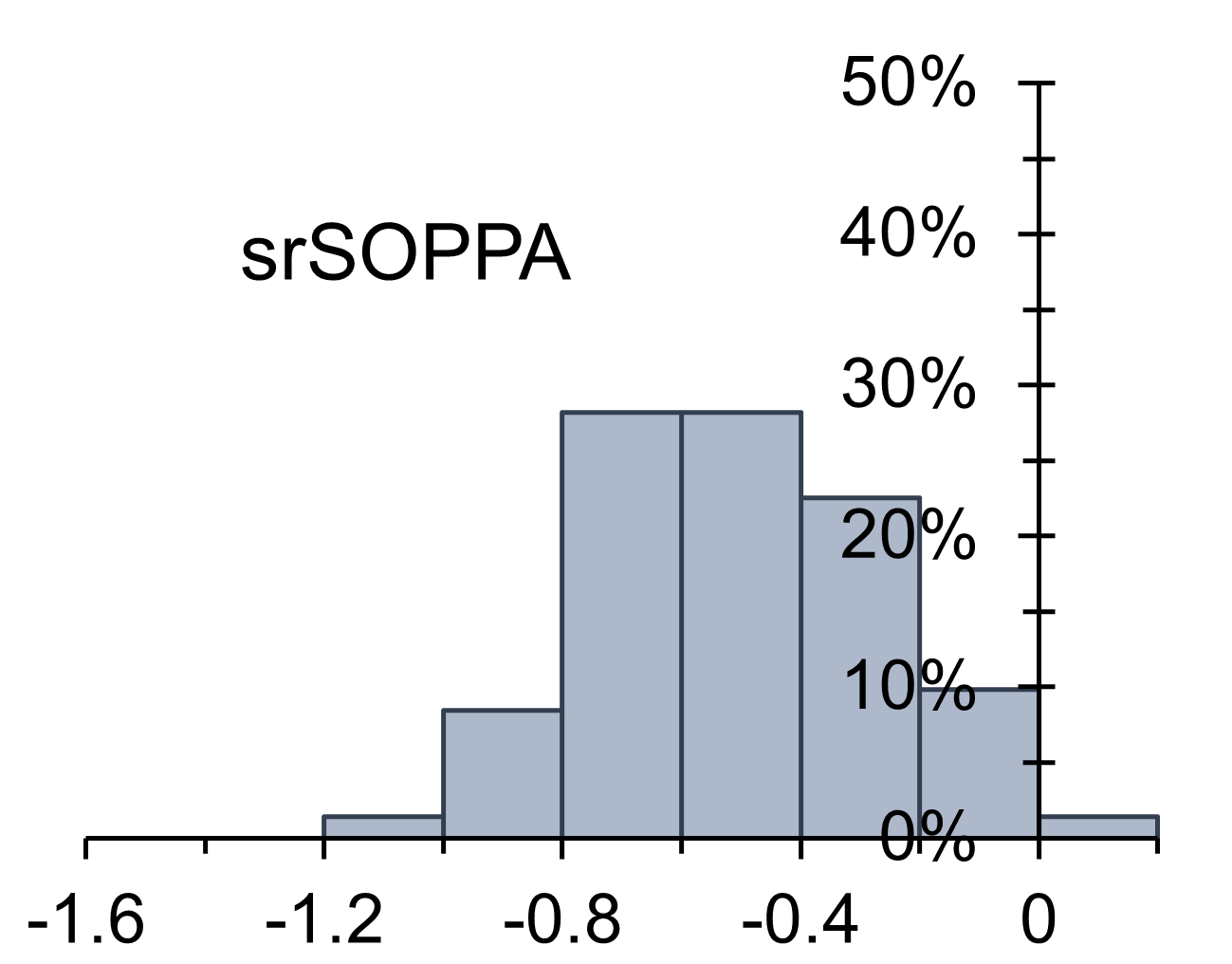}}
    \subfloat{\includegraphics[width=5cm]{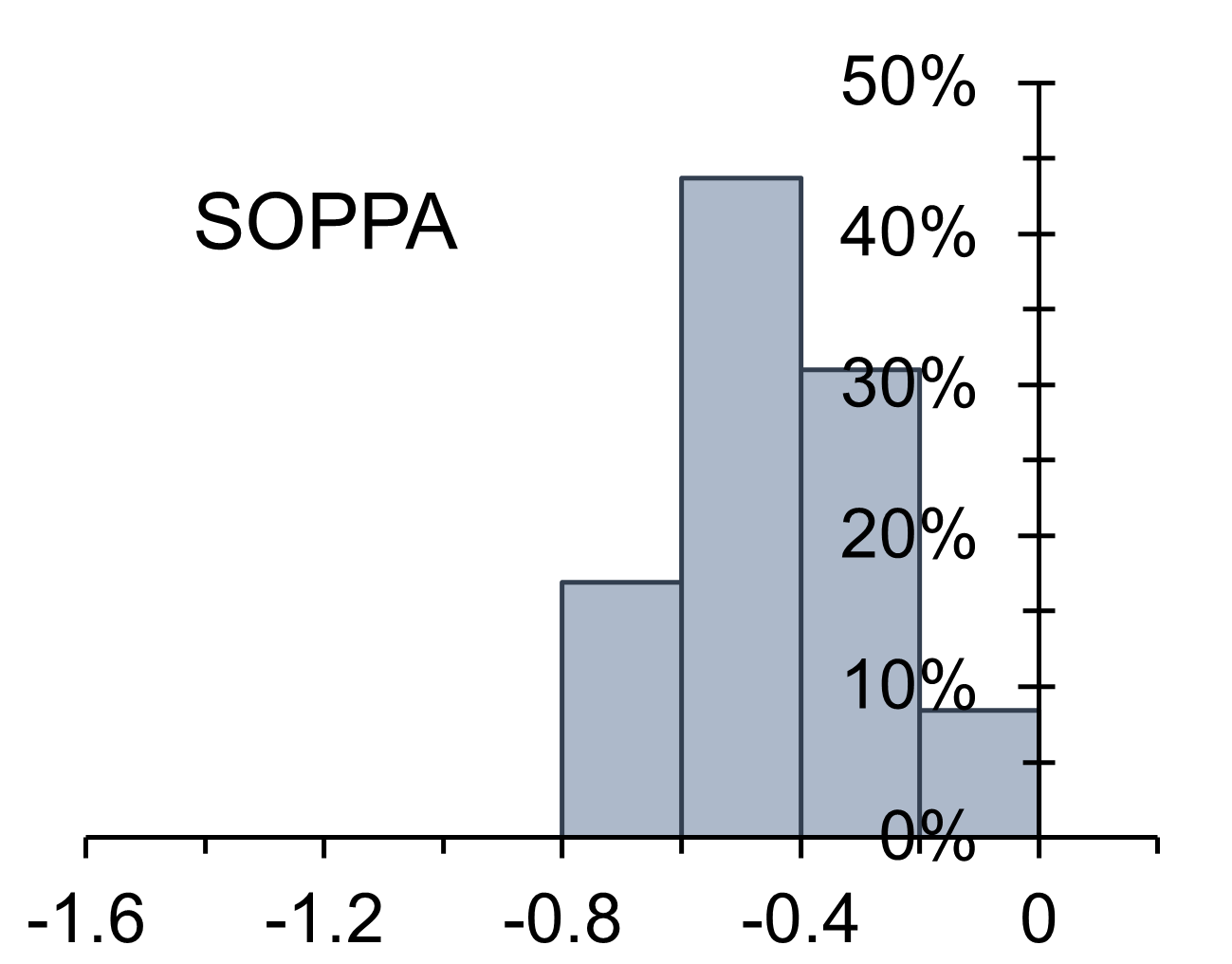}}
    \caption{Histograms (in \%) of the deviations of PBE, SOPPA-srPBE (srSOPPA) and SOPPA vertical triplet excitation energies (in eV) from the CC3 energies in the TZVP basis set.}
    \label{fig:trip_dev}
\end{figure}

\begin{figure}[h!]
    \centering
    \subfloat{\includegraphics[width=5cm]{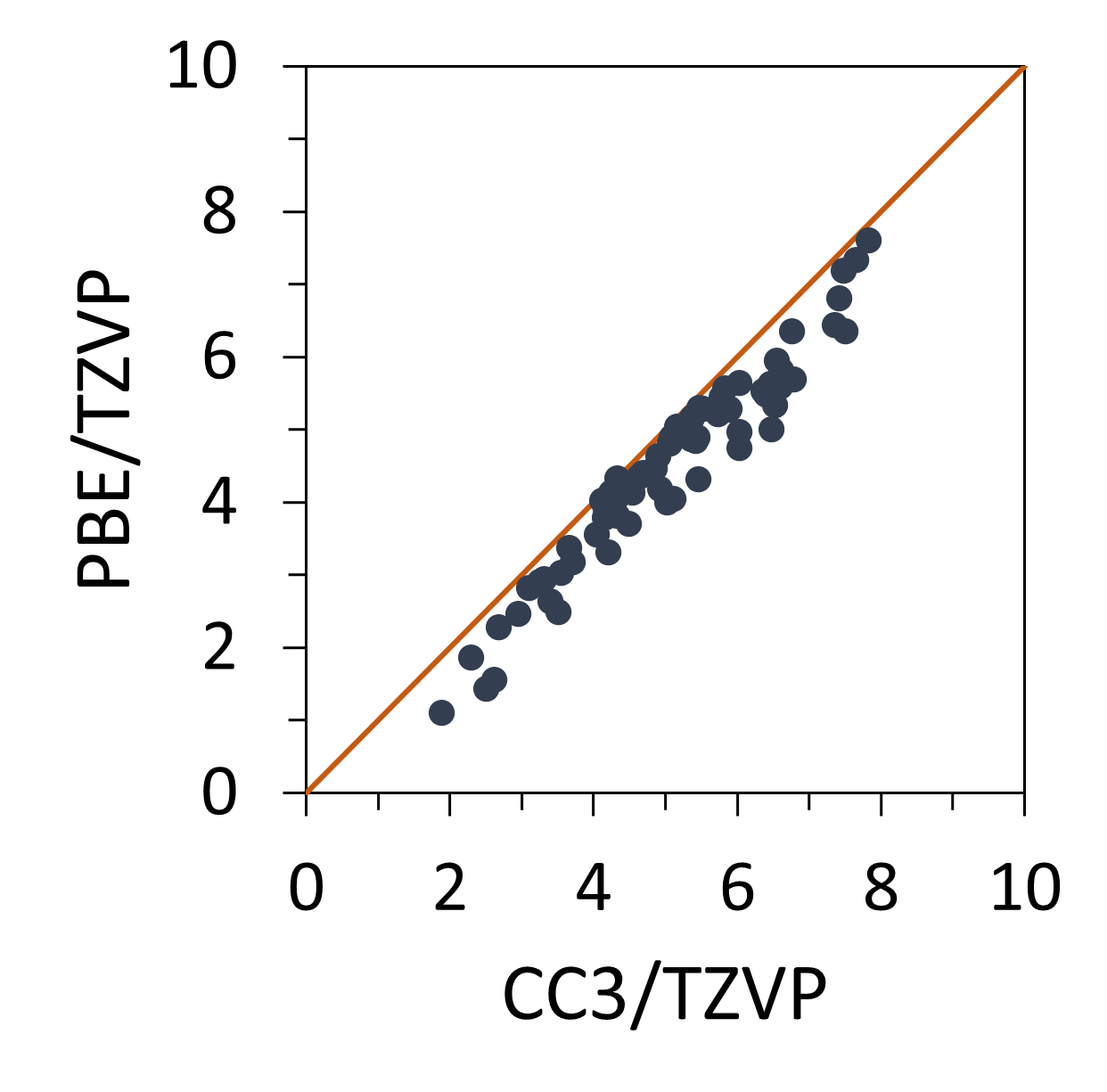}}
    \subfloat{\includegraphics[width=5cm]{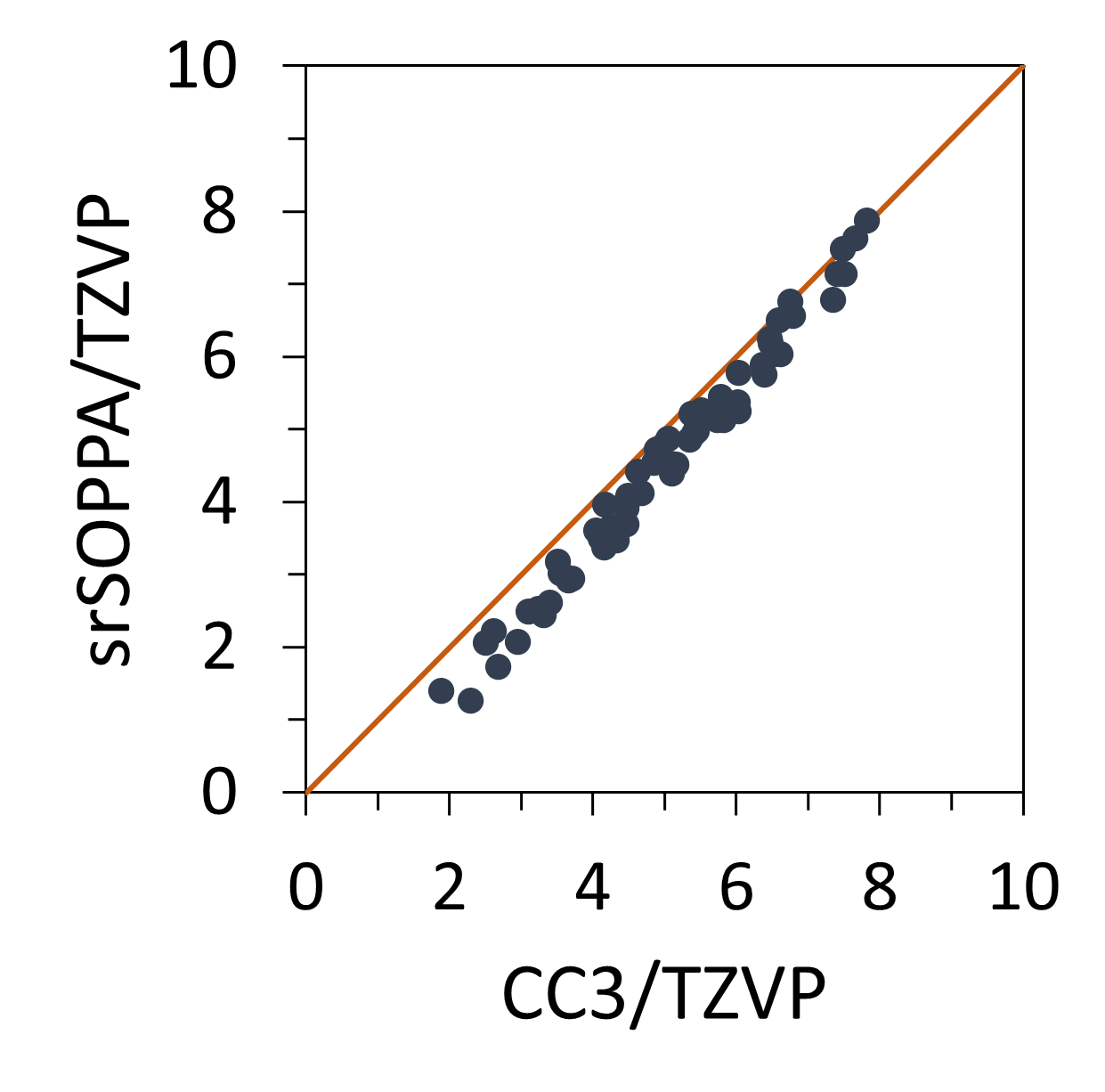}}
    \subfloat{\includegraphics[width=5cm]{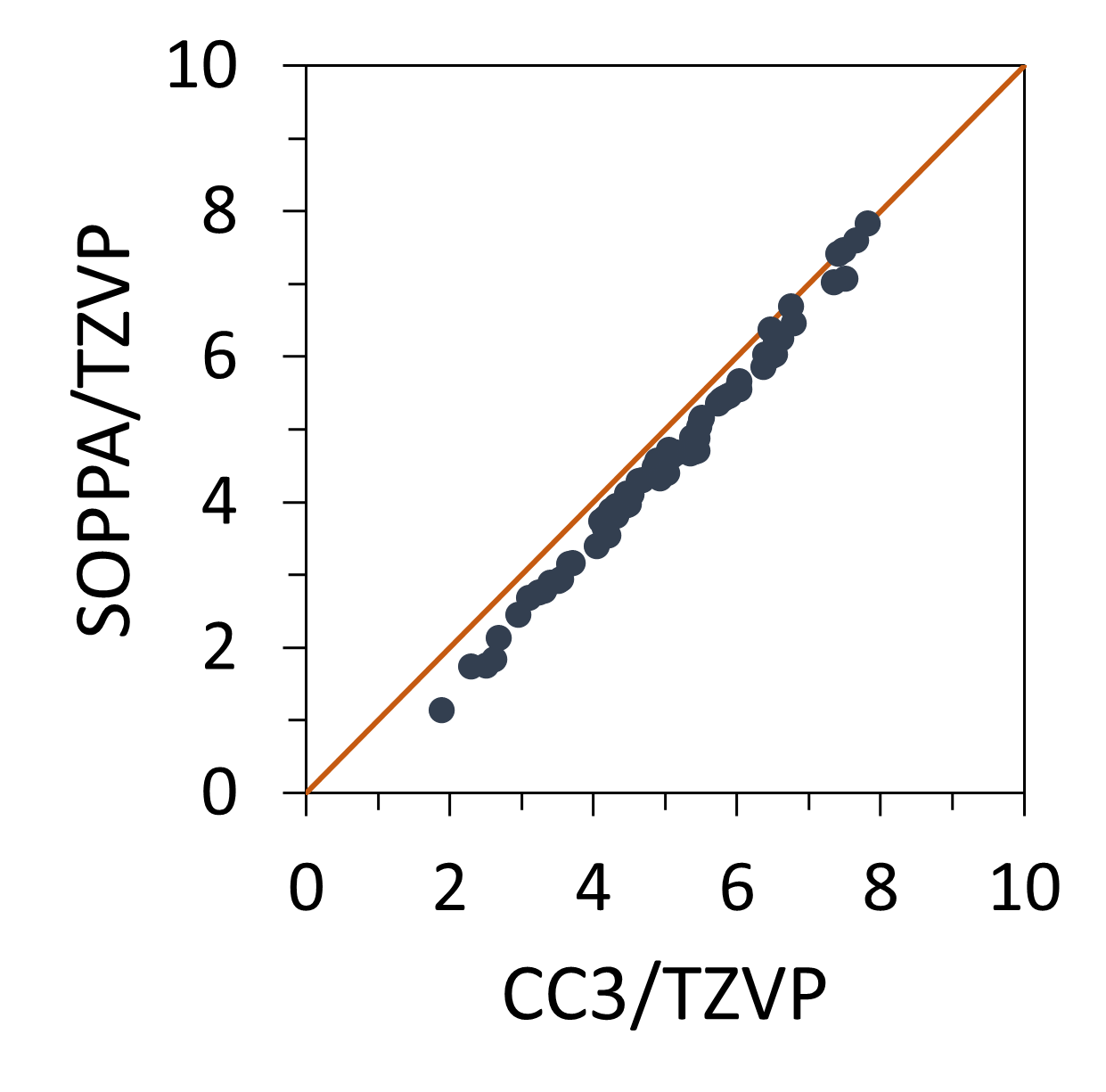}}
    \caption{Correlation plot of the vertical triplet excitation energies (in eV) with PBE, SOPPA-srPBE (srSOPPA) and SOPPA versus CC3 in the TZVP basis set.}
    \label{fig:trip_corr}
\end{figure}

In figure \ref{fig:trip_dev} the deviations from the CC3 results are presented as histograms for the PBE, SOPPA-srPBE, and SOPPA results in the TZVP basis set. The corresponding correlation plots can be seen in figure \ref{fig:trip_corr}. Apart from a few outliers, SOPPA-srPBE underestimates the vertical triplet excitation energies compared to CC3, which is also the case for PBE and SOPPA. From the statistical analysis in table \ref{tab:trip_stat} it can be seen that the mean deviation from CC3 is similar for the three methods SOPPA-srPBE (-0.50 eV), SOPPA (-0.45 eV) and PBE (-0.57 eV) with SOPPA-srPBE inbetween the two other methods and SOPPA the closest to CC3. This is also illustrated in figure \ref{fig:trip_dev} where the columns in the histogram are generally closer to zero for SOPPA than it is for SOPPA-srPBE and PBE. The standard deviations are similarly ranked with the standard deviation for SOPPA-srPBE (0.24 eV) in between the SOPPA (0.17 eV) and PBE (0.34 eV) standard deviations. Again SOPPA performs best between the three methods. This is illustrated in both figure \ref{fig:trip_dev} where the shape of the histograms gets flatter and wider and in figure \ref{fig:trip_corr} where the points in the plots get more dispersed with increasing standard deviation.
\small
\begin{ThreePartTable}
\begin{TableNotes}
    \item[a] Results from \cite{soppa}
    \item[b] Results from \cite{cc32008} 
\end{TableNotes}
\begin{longtable}[h!]{p{2.7cm}p{2.7cm}ccccc}
\caption{Vertical triplet excitation energies (in eV) in the TZVP and aug-cc-pVTZ (aVTZ) basis sets.} \label{tab:trip}

\\ \toprule & & PBE & \multicolumn{2}{c}{srSOPPA} & SOPPA\tnote{a} & CC3\tnote{b}
\\ Molecule & State & TZVP & TZVP & aVTZ & TZVP & TZVP \\ \midrule \endfirsthead

\multicolumn{6}{l}{{\tablename} \thetable{}: \textit{continued.}} \\
\toprule & & PBE & \multicolumn{2}{c}{srSOPPA} & SOPPA\tnote{a} & CC3\tnote{b}
\\ Molecule & State & TZVP & TZVP & aVTZ & TZVP & TZVP \\ \midrule \endhead

\bottomrule \multicolumn{7}{r}{\textit{continued.}} \endfoot 

\bottomrule \insertTableNotes \endlastfoot

Ethene & $1^3$B$_{1u}$ ($\pi$ → $\pi^*$) & 4.25 & 3.68 & 3.68 & 3.95 & 4.48 \\[1.0ex]
E-Butadiene & $1^3$B$_{u }$ ($\pi$ → $\pi^*$) & 2.94 & 2.43 & 2.44 & 2.77 & 3.32 \\
 & $1^3$A$_{g }$ ($\pi$ → $\pi^*$) & 5.03 & 4.51 & 4.48 & 4.68 & 5.17 \\[1.0ex]
all-E-Hexatriene & $1^3$B$_{u }$ ($\pi$ → $\pi^*$) & 2.27 & 1.72 & 1.75 & 2.13 & 2.69 \\
 & $1^3$A$_{g }$ ($\pi$ → $\pi^*$) & 4.05 & 3.60 & 3.60 & 3.80 & 4.32 \\[1.0ex]
all-E-Octatetraene & $1^3$B$_{u }$ ($\pi$ → $\pi^*$) & 1.86 & 1.26 & 1.30 & 1.73 & 2.30 \\
 & $1^3$A$_{g }$ ($\pi$ → $\pi^*$) & 3.36 & 2.91 & 2.92 & 3.14 & 3.67 \\[1.0ex]
Cyclopropene & $1^3$B$_{2 }$ ($\pi$ → $\pi^*$) & 3.81 & 3.47 & 3.44 & 3.89 & 4.34 \\
 & $1^3$B$_{1 }$ ($\sigma$ → $\pi^*$) & 5.80 & 6.03 & 5.89 & 6.24 & 6.62 \\[1.0ex]
Cyclopentadiene & $1^3$B$_{2 }$ ($\pi$ → $\pi^*$) & 2.90 & 2.52 & 2.52 & 2.75 & 3.25 \\
 & $1^3$A$_{1 }$ ($\pi$ → $\pi^*$) & 4.89 & 4.52 & 4.48 & 4.63 & 5.09 \\[1.0ex]
Norbornadiene & $1^3$A$_{2 }$ ($\pi$ → $\pi^*$) & 3.17 & 2.94 & 2.91 & 3.16 & 3.72 \\
 & $1^3$B$_{2 }$ ($\pi$ → $\pi^*$) & 3.78 & 3.37 & 3.35 & 3.64 & 4.16 \\[1.0ex]
Benzene & $1^3$B$_{1u}$ ($\pi$ → $\pi^*$) & 4.01 & 3.49 & 3.48 & 3.73 & 4.12 \\
 & $1^3$E$_{1u}$ ($\pi$ → $\pi^*$) & 4.63 & 4.71 & 4.63 & 4.56 & 4.90 \\
 & $1^3$B$_{2u}$ ($\pi$ → $\pi^*$) & 4.96 & 5.24 & 5.09 & 5.66 & 6.04 \\
 & $1^3$E$_{2g}$ ($\pi$ → $\pi^*$) & 7.18 & 7.48 & 7.41 & 7.46 & 7.49 \\[1.0ex]
Naphthalene & $1^3$B$_{2u}$ ($\pi$ → $\pi^*$) & 2.81 & 2.48 & 2.47 & 2.68 & 3.11 \\
 & $1^3$B$_{3u}$ ($\pi$ → $\pi^*$) & 3.83 & 3.95 & 3.87 & 3.78 & 4.18 \\
 & $1^3$B$_{1g}$ ($\pi$ → $\pi^*$) & 4.24 & 4.00 & 3.95 & 4.04 & 4.47 \\
 & $2^3$B$_{2u}$ ($\pi$ → $\pi^*$) & 4.35 & 4.41 & 4.34 & 4.29 & 4.64 \\
 & $2^3$B$_{3u}$ ($\pi$ → $\pi^*$) & 4.04 & 4.38 & 4.25 & 4.66 & 5.11 \\
 & $1^3$A$_{g }$ ($\pi$ → $\pi^*$) & 5.28 & 5.25 & 5.17 & 5.15 & 5.52 \\
 & $2^3$B$_{1g}$ ($\pi$ → $\pi^*$) & 5.00 & 6.17 & 5.89 & 6.00 & 6.48 \\
 & $2^3$A$_{g }$ ($\pi$ → $\pi^*$) & 5.62 & 6.23 & 6.02 & 6.37 & 6.47 \\
 & $3^3$A$_{g }$ ($\pi$ → $\pi^*$) & 5.68 & 6.55 & 6.44 & 6.46 & 6.79 \\
 & $3^3$B$_{1g}$ ($\pi$ → $\pi^*$) & 6.34 & 6.75 & 6.62 & 6.69 & 6.76 \\[1.0ex]
Furan & $1^3$B$_{2 }$ ($\pi$ → $\pi^*$) & 3.91 & 3.54 & 3.49 & 3.77 & 4.17 \\
 & $1^3$A$_{1 }$ ($\pi$ → $\pi^*$) & 5.29 & 5.10 & 5.03 & 5.03 & 5.48 \\[1.0ex]
Pyrrole & $1^3$B$_{2 }$ ($\pi$ → $\pi^*$) & 4.24 & 3.90 & 3.86 & 4.11 & 4.48 \\
 & $1^3$A$_{1 }$ ($\pi$ → $\pi^*$) & 5.28 & 5.21 & 5.13 & 5.13 & 5.51 \\[1.0ex]
Imidazole & $1^3$A$'$ ($\pi$ → $\pi^*$) & 4.39 & 4.11 & 4.07 & 4.30 & 4.69 \\
 & $2^3$A$'$ ($\pi$ → $\pi^*$) & 5.44 & 5.44 & 5.30 & 5.40 & 5.79 \\
 & $1^3$A$''$ (n → $\pi^*$) & 5.52 & 5.88 & 5.83 & 5.85 & 6.37 \\
 & $3^3$A$'$ ($\pi$ → $\pi^*$) & 5.94 & 6.05 & 5.93 & 6.21 & 6.55 \\
 & $4^3$A$'$ ($\pi$ → $\pi^*$) & 6.80 & 7.13 & 6.85 & 7.41 & 7.42 \\
 & $2^3$A$''$ (n → $\pi^*$) & 6.34 & 7.13 & 7.00 & 7.07 & 7.51 \\[1.0ex]
Pyridine & $1^3$A$_{1 }$ ($\pi$ → $\pi^*$) & 4.13 & 3.64 & 3.63 & 3.88 & 4.25 \\
 & $1^3$B$_{1 }$ (n → $\pi^*$) & 3.69 & 4.08 & 4.04 & 3.97 & 4.5 \\
 & $1^3$B$_{2 }$ ($\pi$ → $\pi^*$) & 4.45 & 4.53 & 4.41 & 4.48 & 4.86 \\
 & $2^3$A$_{1 }$ ($\pi$ → $\pi^*$) & 4.80 & 4.86 & 4.78 & 4.71 & 5.06 \\
 & $1^3$A$_{2 }$ (n → $\pi^*$) & 4.31 & 5.06 & 4.99 & 4.87 & 5.46 \\
 & $2^3$B$_{2 }$ ($\pi$ → $\pi^*$) & 5.48 & 5.75 & 5.63 & 6.03 & 6.4 \\
 & $3^3$B$_{2 }$ ($\pi$ → $\pi^*$) & 7.6 & 7.87 & 7.75 & 7.83 & 7.83 \\
 & $3^3$A$_{1 }$ ($\pi$ → $\pi^*$) & 7.33 & 7.62 & 7.56 & 7.59 & 7.66 \\[1.0ex]
s-Tetrazine & $1^3$B$_{3u}$ (n → $\pi^*$) & 1.09 & 1.39 & 1.40 & 1.13 & 1.89 \\
 & $1^3$A$_{u }$ (n → $\pi^*$) & 2.48 & 3.17 & 3.17 & 2.91 & 3.52 \\
 & $1^3$B$_{1g}$ (n → $\pi^*$) & 3.3 & 3.61 & 3.62 & 3.53 & 4.21 \\
 & $1^3$B$_{1u}$ ($\pi$ → $\pi^*$) & 4.32 & 3.65 & 3.65 & 3.94 & 4.33 \\
 & $1^3$B$_{2u}$ ($\pi$ → $\pi^*$) & 4.13 & 4.06 & 3.93 & 4.10 & 4.54 \\
 & $1^3$B$_{2g}$ (n → $\pi^*$) & 4.17 & 4.55 & 4.54 & 4.32 & 4.93 \\
 & $2^3$A$_{u }$ (n → $\pi^*$) & 3.99 & 4.52 & 4.49 & 4.40 & 5.03 \\
 & $2^3$B$_{1u}$ ($\pi$ → $\pi^*$) & 5.15 & 5.21 & 5.13 & 4.88 & 5.38 \\
 & $2^3$B$_{2g}$ (n → $\pi^*$) & 4.74 & 5.77 & 5.74 & 5.55 & 6.04 \\
 & $2^3$B$_{1g}$ (n → $\pi^*$) & 5.59 & 6.49 & 6.45 & 6.23 & 6.6 \\
 & $2^3$B$_{3u}$ (n → $\pi^*$) & 5.33 & 6.03 & 5.99 & 6.02 & 6.53 \\
 & $2^3$B$_{2u}$ ($\pi$ → $\pi^*$) & 6.43 & 6.77 & 6.69 & 7.02 & 7.36 \\[1.0ex]
Formaldehyde & $1^3$A$_{2 }$ ($n$ → $\pi^*$) & 3.02 & 3.01 & 2.98 & 2.93 & 3.55 \\
 & $1^3$A$_{1 }$ ($\pi$ → $\pi^*$) & 5.56 & 5.12 & 5.12 & 5.42 & 5.83 \\[1.0ex]
Acetone & $1^3$A$_{2 }$ (n → $\pi^*$) & 3.55 & 3.60 & 3.59 & 3.39 & 4.05 \\
 & $1^3$A$_{1 }$ ($\pi$ → $\pi^*$) & 5.63 & 5.37 & 5.38 & 5.60 & 6.03 \\[1.0ex]
p-Benzoquinone & $1^3$B$_{1g}$ (n → $\pi^*$) & 1.42 & 2.05 & 2.04 & 1.74 & 2.51 \\
 & $1^3$A$_{u }$ (n → $\pi^*$) & 1.55 & 2.22 & 2.22 & 1.83 & 2.62 \\
 & $1^3$B$_{1u}$ ($\pi$ → $\pi^*$) & 2.46 & 2.07 & 2.13 & 2.44 & 2.96 \\
 & $1^3$B$_{3g}$ ($\pi$ → $\pi^*$) & 2.63 & 2.61 & 2.6 & 2.89 & 3.41 \\[1.0ex]
Formamide & $1^3$A$''$ (n → $\pi^*$) & 4.86 & 4.85 & 4.78 & 4.66 & 5.36 \\
 & $1^3$A$'$ ($\pi$ → $\pi^*$) & 5.21 & 5.12 & 5.07 & 5.35 & 5.74 \\[1.0ex]
Acetamide & $1^3$A$''$ (n → $\pi^*$) & 4.84 & 4.94 & 4.88 & 4.69 & 5.42 \\
 & $1^3$A$'$ ($\pi$ → $\pi^*$) & 5.27 & 5.27 & 5.24 & 5.45 & 5.88 \\[1.0ex]
Propanamide & $1^3$A$''$ (n → $\pi^*$) & 4.88 & 4.97 & 4.90 & 4.70 & 5.45 \\
 & $1^3$A$'$ ($\pi$ → $\pi^*$) & 5.28 & 5.30 & 5.26 & 5.46 & 5.90 \\

\end{longtable} \end{ThreePartTable} \normalsize
\begin{table}[h!]
\centering 
\caption{Deviations of the vertical triplet excitation energies (in eV) in the TZVP basis set from the CC3/TZVP results.}
\label{tab:trip_stat}
\small
\begin{tabular}{cccc} \toprule 
& PBE & srSOPPA & SOPPA \\

\midrule   &  & all &  \\
Count & 71 & 71 & 71 \\
Mean & -0.57 & -0.50 & -0.45 \\
Abs. mean & 0.57 & 0.50 & 0.45 \\
Std. dev. & 0.34 & 0.24 & 0.17 \\
Maximum (+) & - & 0.04 & - \\
Maximum (-) & 1.48 & 1.04 & 0.79 \\
\midrule   &  & ($\pi$ → $\pi^*$) &  \\
Count & 51 & 51 & 51 \\
Mean & -0.44 & -0.53 & -0.39 \\
Abs. mean & 0.44 & 0.53 & 0.39 \\
Std. dev. & 0.30 & 0.27 & 0.14 \\
Maximum (+) & - & 0.04 & - \\
Maximum (-) & 1.48 & 1.04 & 0.57 \\
\midrule   &  & (n → $\pi^*$) &  \\
Count & 19 & 19 & 19 \\
Mean & -0.89 & -0.43 & -0.62 \\
Abs. mean & 0.89 & 0.43 & 0.62 \\
Std. dev. & 0.25 & 0.11 & 0.12 \\
Maximum (-) & 1.30 & 0.60 & 0.79 \\

\bottomrule \end{tabular}\end{table} \normalsize


From figure \ref{fig:trip_dev} it can be seen that SOPPA has the fewest outliers both above and the mean deviation. SOPPA-srPBE has more outliers primarily below the mean deviation and PBE has by far the most outliers. Investigating figure \ref{fig:trip_corr} it can visually be seen that both SOPPA-srPBE and SOPPA deviate more from the CC3 results for lower excitation energies and that the general agreement with CC3 is better for excitation energies above $\sim5$ eV. PBE performs equally poorly across all excitations. For SOPPA-srPBE the largest outliers above the mean deviation from the CC3 results are the $1^3$E$_{2g}$ state in benzene, $3^3$B$_{1g}$ state in naphthalene, and $3^3$B$_2$ and $3^3$A$_1$ states in pyridine. For these states SOPPA-srPBE is in very good agreement with the CC3 results with an absolute deviation of less than 0.1 eV. For the same states SOPPA is also in good agreement with the CC3 results \cite{soppa}, in addition SOPPA is in good agreement with CC3 for the $2^3$A$_g$ state in naphthalene and $4^3$A' state in imidazole. For both methods these states are some of the higher excitation energies and all of them are larger than $5$ eV.

The largest outliers below the mean deviation from the CC3 results for SOPPA-srPBE are the $1^3$B$_u$ states of E-butadiene, all-E-hexatriene and all-E-octatetraene, $1^3$B$_2$ state of cyclopropene, and $1^3$B$_{1u}$ of p-benzoquinone. The worst states for PBE are the $2^3$B$_{1g}$ state of naphthalene, and $2^3$B$_{2g}$ and $2^3$B$_{3u}$ states in s-tetrazine. While the states performing well for SOPPA and SOPPA-srPBE are similar, the largest deviations below the mean deviations for SOPPA are primarily $n \rightarrow \pi^*$ transitions \cite{soppa}, where for SOPPA-srPBE all the worst states are $\pi \rightarrow \pi^*$ transitions. Comparing the $n \rightarrow \pi^*$ transitions of the SOPPA and SOPPA-srPBE in table \ref{tab:trip}, SOPPA-srPBE outperforms SOPPA for most of the states despite SOPPA performing better overall. To further examine this, the statistical analysis in table \ref{tab:trip_stat} has been separated into $\pi \rightarrow \pi^*$ and $n \rightarrow \pi^*$ transitions. For the $\pi \rightarrow \pi^*$ transitions SOPPA-srPBE is worse than SOPPA with larger absolute mean deviation and a larger standard deviation. SOPPA-srPBE and PBE are closer with SOPPA-srPBE having a slightly larger mean deviation and a slightly smaller standard deviation. However for the $n \rightarrow \pi*$ transitions SOPPA-srPBE has the smallest absolute mean deviation and SOPPA-srPBE and SOPPA have similar standard deviations. PBE is performing far worse and has an over twice as large mean and standard deviation as SOPPA.

Comparing with B3LYP for the same benchmarking set, the mean deviation for SOPPA-srPBE is $-0.51\pm 0.24$ eV where it's $-0.48\pm 0.51$ eV for B3LYP \cite{dft2008}. The mean deviation are comparable, but the consistency of the SOPPA-srPBE results is better as indicated by the standard deviation which is about half of the standard deviation for B3LYP.

\subsection{Basis set analysis}
The effect of adding the extra diffuse functions in the aug-cc-pVTZ basis set has previously been studied for vertical singlet excitation energies in Thiel benchmarking set for an array of methods: RPA(D), SOPPA, SOPPA(CCSD), CC2, CC3, and CASPT2 \cite{soppa,cc32010,basiseffectCAS}. It has been observed in these studies, that adding the extra diffuse functions decreases the vertical singlet excitation energies by on average $-0.24\pm0.17$ eV for SOPPA, $-0.26\pm0.17$ eV for SOPPA(CCSD), $-0.22\pm0.29$ eV for CC2, $-0.18\pm0.25$ eV for CC3, and $-0.11\pm0.22$ eV for CASPT2. The exception is RPA(D), where the energy increases for a few of the states.
\begin{table}[h!]
\centering
\caption{Deviations in the vertical excitation energies of SOPPA-srPBE in the aug-cc-pVTZ basis set from the TZVP results.} \label{tab:stat_basis} \small \begin{tabular}{ccc} 
\toprule & Singlet & Triplet \\ \midrule
Count & 132 & 71 \\
Mean & -0.22 & -0.06 \\
Abs. mean & 0.22 & 0.06 \\
Std. dev. & 0.25 & 0.06 \\
Maximum (+) & - & 0.06 \\
Maximum (-) & 1.81 & 0.28 \\
\bottomrule\end{tabular}\end{table}

\begin{figure}[h!]
    \centering
    \subfloat{\includegraphics[width=5cm]{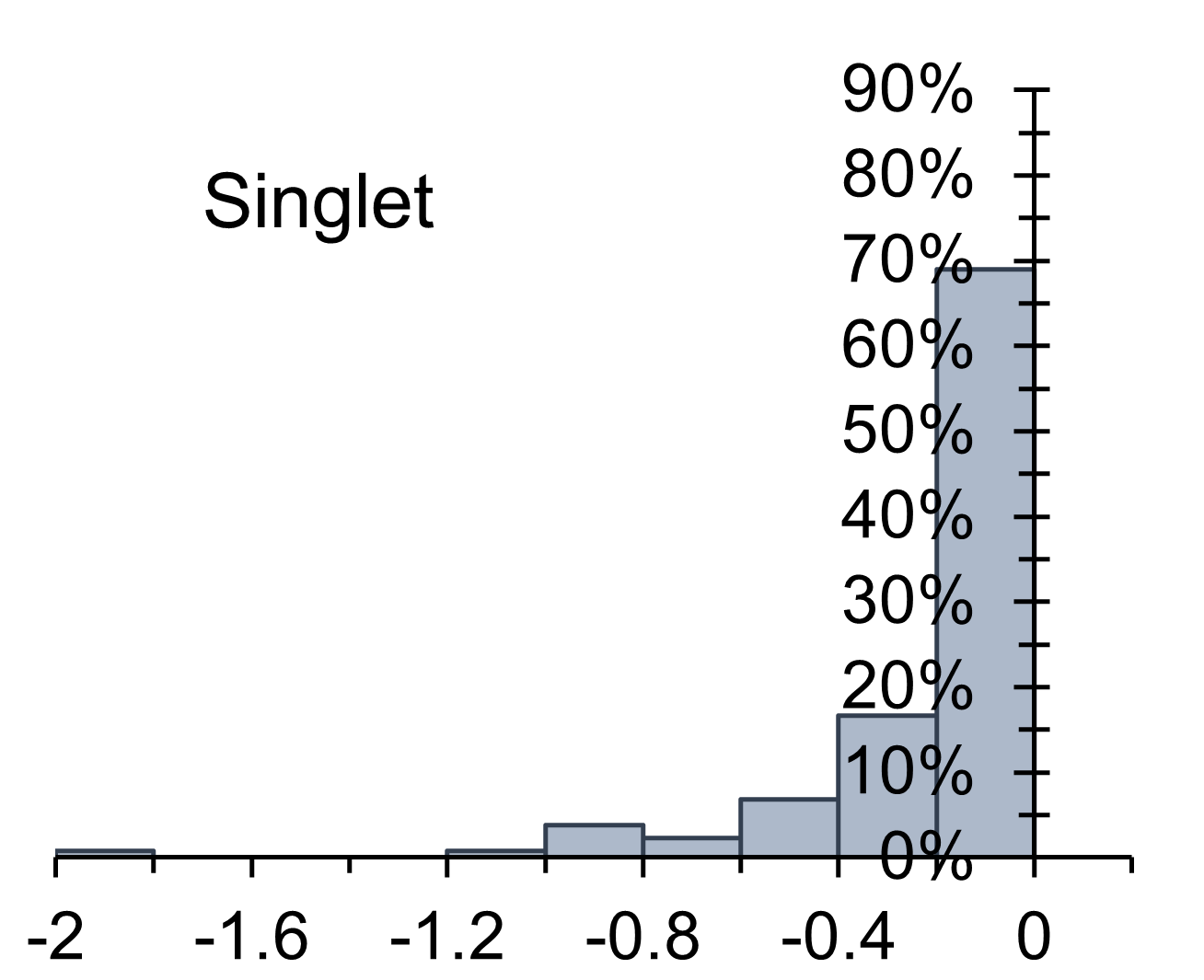}}
    \subfloat{\includegraphics[width=5cm]{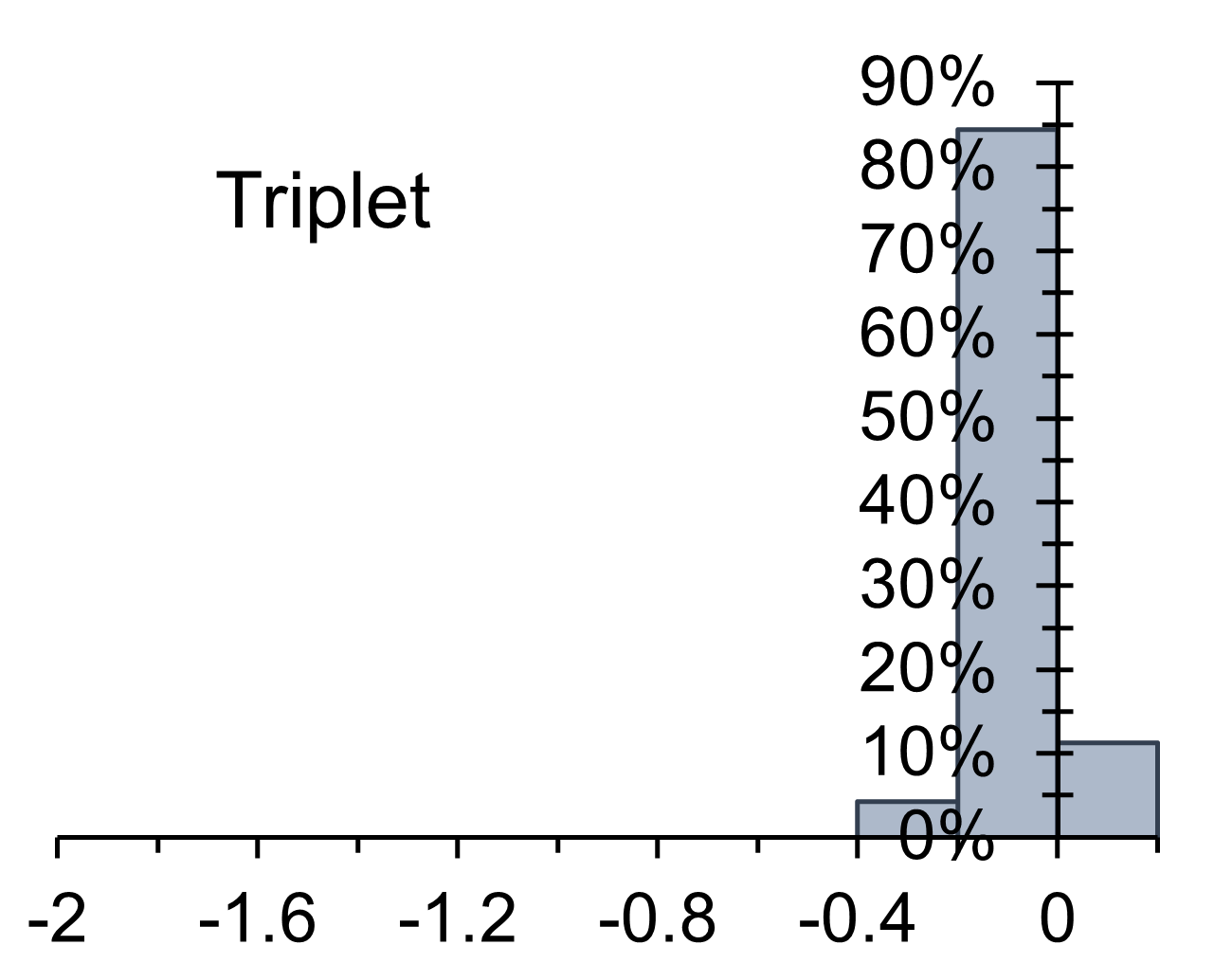}}
    \caption{Histograms (in \%) of the deviations of SOPPA-srPBE/aug-cc-pVTZ from SOPPA-srPBE/TZVP vertical singlet and triplet excitation energies (in eV).}
    \label{fig:basis_dev}
\end{figure}

In figure \ref{fig:basis_dev} it can be seen that the addition of the diffuse basis functions reduces the vertical excitation energies of SOPPA-srPBE for all states, with the exception of about 10\% of the triplet states. A statistical analysis of the deviations from the TZVP results is collected in table \ref{tab:stat_basis}. For the singlet states the mean shift of $-0.22$ eV is similar to but slightly smaller than the previously recorded shifts of other SOPPA based methods. For the triplet states the mean shift of $-0.06$ eV is smaller than any of the previously reported shifts. However it is important to note that all the states previously investigated were singlet states.

A similar trend is present in the standard deviations. For the singlet excited states the standard deviation is $0.25$ eV which is comparable to the standard deviations found in the previous studies. For the triplet excited states the standard deviation is $0.06$ eV, which is much smaller than any of the previously recorded standard deviations \cite{soppa,cc32010,basiseffectCAS}.

\begin{figure}[h!]
    \centering
    \subfloat{\includegraphics[width=5cm]{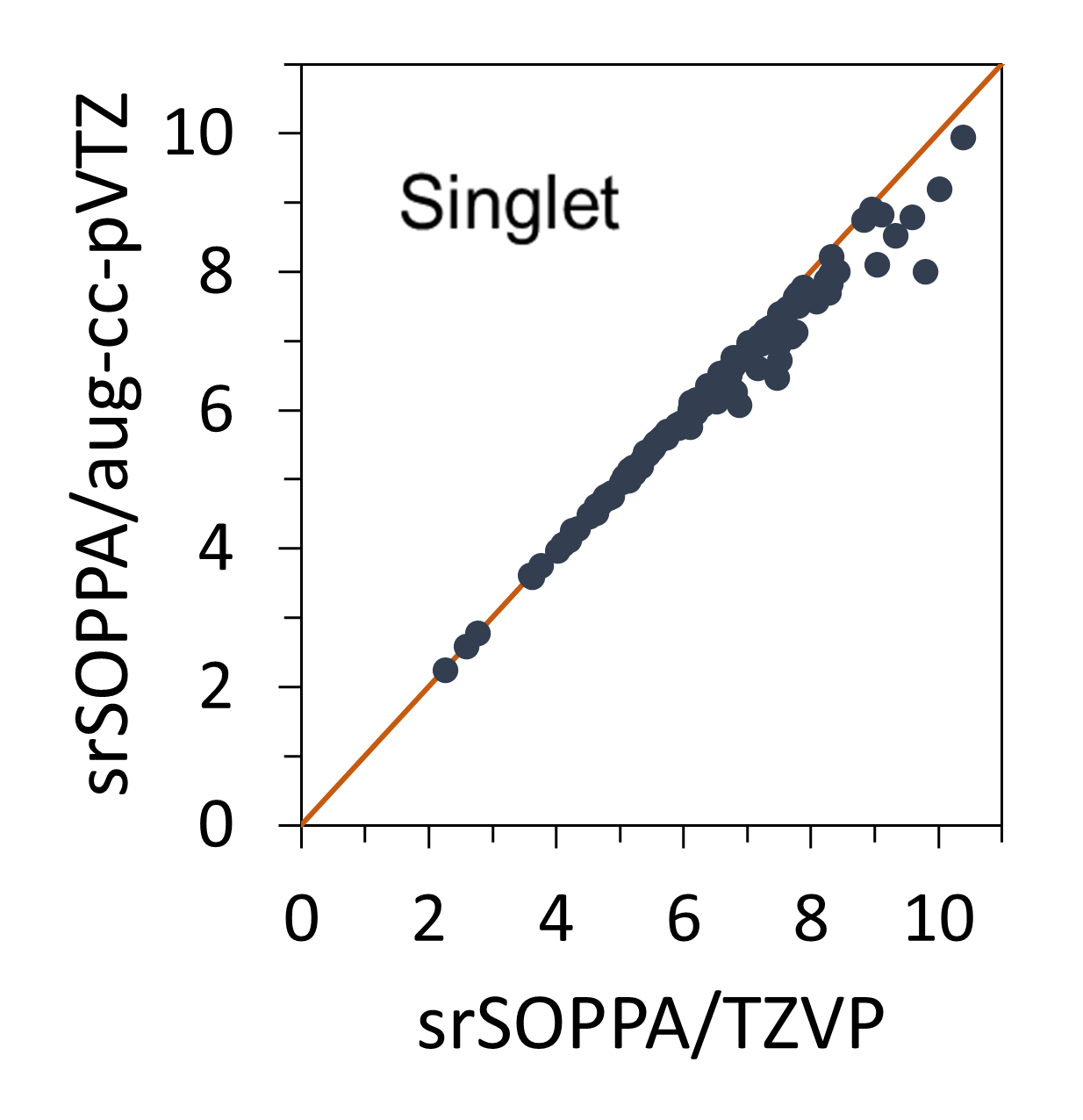}}
    \subfloat{\includegraphics[width=5cm]{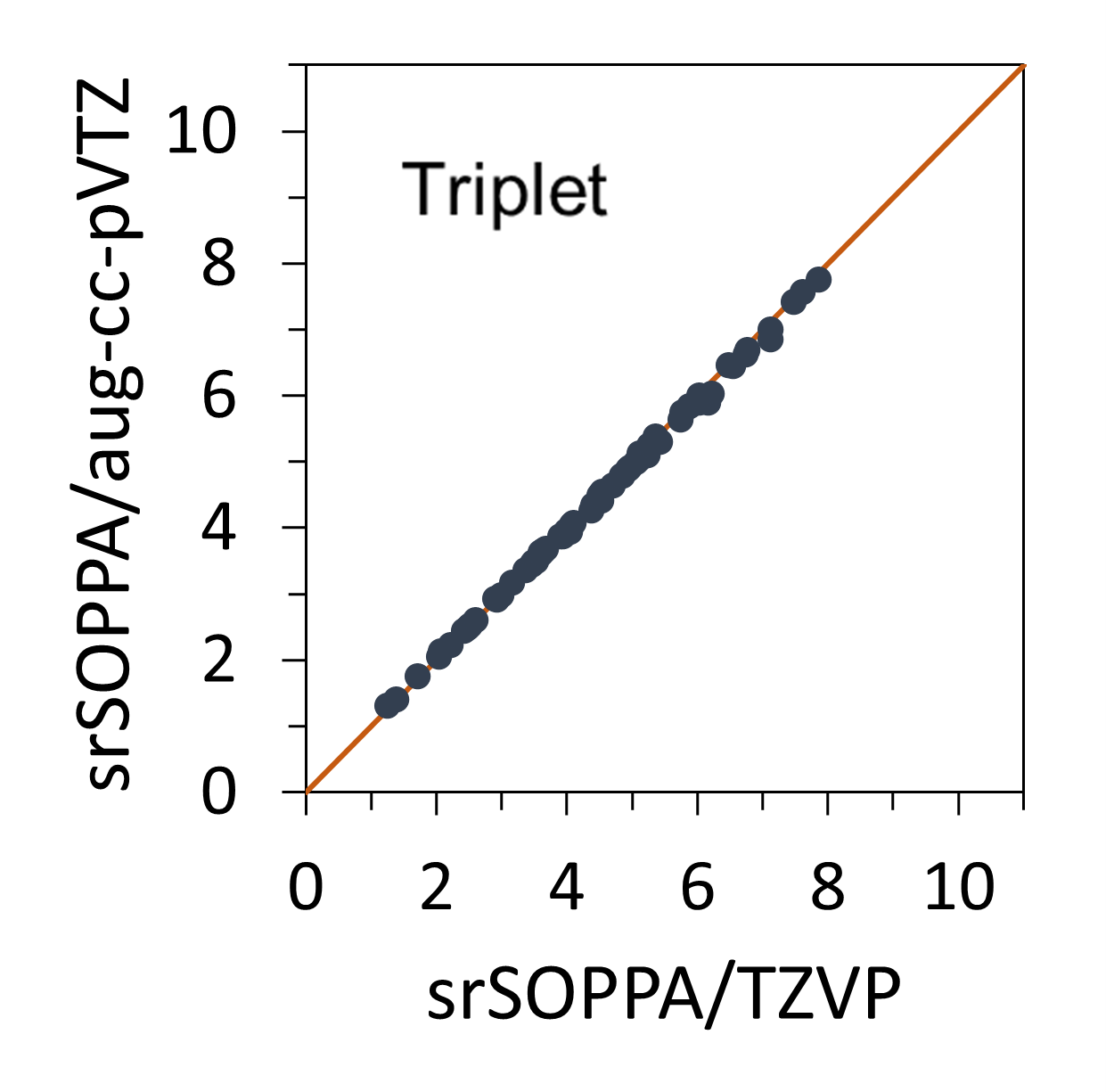}}
    \caption{Correlation plot of the vertical excitation energies (in eV) of SOPPA-srPBE in the aug-cc-pVTZ basis set versus the TZVP basis set results.}
    \label{fig:basis_corr}
\end{figure}

From the correlation plot in figure \ref{fig:basis_corr} it can be seen that the effect of the basis set is larger for the larger excitation energies for the singlet states. This is also what leads to the trail of larger deviations in the histogram in figure \ref{fig:basis_dev}. This effect has also previously been observed for singlet excitations calculated with other SOPPA based methods \cite{soppa}. This effect is not present for the triplet excitations.

\section{Conclusions}
In this paper the performance of SOPPA-srPBE has been investigated for the calculation of vertical electronic excitation energies. SOPPA-srPBE calculations were carried out for the 132 singlets states and 71 triplet states in a previously proposed benchmarking set of medium sized organic molecules in both the TZVP and aug-cc-pVTZ basis sets. The results were compared to corresponding PBE, SOPPA and CC3 results. Additionally the effect of adding the extra diffuse basis functions from the larger aug-cc-pVTZ basis set was investigated. 

In the calculations of singlet excitation energies, the SOPPA-srPBE both under and overestimates the energies, which results in a smaller absolute mean deviation but larger standard deviation. When adding diffuse functions the standard deviation becomes the same for the two method, showing that with larger basis sets SOPPA-srPBE is even more reliable. In general SOPPA-srPBE performs better than SOPPA overall with an average difference of $0.30\pm0.37$ eV for SOPPA-srPBE versus $0.50\pm0.28$ eV for SOPPA with the TZVP, and with $0.24\pm0.29$ eV for SOPPA-srPBE versus $0.50\pm0.28$ eV for SOPPA with the aug-cc-pVTZ basis set.

When calculating triplet excitation energies, it can be concluded that overall SOPPA-srPBE performs slightly worse than SOPPA, but better than PBE. SOPPA-srPBE underestimates, with one exception, the CC3 triplet excitation energies on average by $0.50\pm0.24$ eV. In comparison, SOPPA also systematically underestimates the CC3 triplet excitation energies, but the average difference is smaller ($0.45\pm0.17$ eV). For $n\rightarrow\pi^*$ transitions SOPPA-srPBE performs better than SOPPA overall with an average difference of $0.43\pm0.11$ eV for SOPPA-srPBE versus $0.62\pm0.12$ eV for SOPPA. However it should be noted that only 19 of the triplet states in this study are $n\rightarrow\pi^*$. 

The effect of adding extra diffuse basis functions in the aug-cc-pVTZ basis set overall results in a decrease of the vertical excitation energies, but about 10\% of the investigated states increase in energy in the aug-cc-pVTZ basis set. The mean shift for singlet excited states is considerably larger ($-0.22$ eV) than for triplet excited states ($-0.06$ eV)

To summarise SOPPA-srPBE outperforms SOPPA in the calculation of vertical singlet excitation energies, but not when calculating vertical triplet excitation energies.  For the singlet states SOPPA-srPBE has been found to have a smaller mean absolute deviation than SOPPA with both basis sets, and have similar standard deviation using the augmented basis. For the triplet states SOPPA-srPBE has been found to perform slightly worse than SOPPA in terms of both mean absolute deviation and standard deviation. 
Finally, SOPPA-srPBE outperforms PBE for both types of excitation energies. In conclusion based on this study, we recommend to prefer SOPPA-srPBE over SOPPA for singlet excitation eneriges and vice versa for triplet excitation energies.

\printbibliography
\end{document}